\documentclass[3p,twocolumn,10pt,sort&compress]{elsarticle}
\usepackage{graphicx}
\usepackage[american]{babel}
\usepackage[utf8]{inputenc}
\usepackage[T1]{fontenc}
\usepackage{latexsym,amsmath,amssymb}

\usepackage[numbers]{natbib}
\usepackage{float}
\usepackage{placeins}
\usepackage{url}

\allowdisplaybreaks
\usepackage[sc]{mathpazo}
\usepackage{caption}
\usepackage[ruled,vlined,noend]{algorithm2e}

\usepackage{booktabs}
\usepackage{tabularx}
\usepackage{threeparttable}
\usepackage{siunitx}
\usepackage[gen]{eurosym}

\usepackage{color}

\usepackage[hidelinks]{hyperref}
\usepackage{cleveref}

\usepackage{adjustbox}

\newcommand{\fref}[1]{Figure~\ref{#1}}
\newcommand{\tref}[1]{Table~\ref{#1}}
\renewcommand{\eqref}[1]{Equation~(\ref{#1})}

\usepackage{nomencl}
\usepackage{framed}
\makenomenclature

\usepackage{doi}

\begin{document}

\begin{frontmatter}

\title{The role of hydro power, storage and transmission in the decarbonization of the Chinese power system}

\journal{Applied Energy}

\author[label1]{Hailiang Liu}
\ead{HLL@eng.au.dk}
\author[label2]{Tom Brown}
\author[label1]{Gorm Bruun Andresen}
\author[label1]{David P. Schlachtberger}
\author[label1]{Martin Greiner}
\address[label1]{Department of Engineering, Aarhus University, Inge Lehmanns Gade 10, 8000 Aarhus C,  Denmark}
\address[label2]{Institute for Automation and Applied Informatics, Karlsruhe Institute of Technology, 76344 Eggenstein-Leopoldshafen, Germany}

\begin{abstract}

Deep decarbonization of the electricity sector can be provided by a high penetration of renewable sources such as wind, solar PV and hydro power. Flexibility from hydro and storage complements the high temporal variability of wind and solar, and transmission infrastructure helps the power balancing by moving electricity in the spatial dimension. We study cost-optimal highly-renewable Chinese power systems under
ambitious CO$ _2 $ emission reduction targets, by deploying a 31-node hourly-resolved techno-economic optimization model supported by a validated weather-converted 38-year-long renewable power generation and electricity demand dataset. With a new realistic reservoir hydro model, we find that if CO$_2$ emission reduction goes beyond 70\%, storage facilities such as hydro, battery and hydrogen become necessary for a moderate system cost. Numerical results show that these flexibility components can lower renewable curtailment by two thirds, allow higher solar PV share by a factor of two and contribute to covering summer cooling demand. We show that expanding unidirectional high-voltage DC lines on top of the regional inter-connections is technically sufficient and more economical than ultra-high-voltage-AC-connected "One-Net" grid. Finally, constraining transmission volume from the optimum by up to 25\% does not push total costs much higher, while the significant need for battery storage remains even with abundant interconnectivity.

\end{abstract}


\end{frontmatter}

\clearpage

\section{Introduction}

The Chinese power sector accounts for almost half of the country's annual CO$_2$ emissions, which is expected to reach 12 Gt in 2020. To honor its own pledge as part of the global effort in curbing climate change and also improve regional air quality, China is in the process of decarbonization by transforming the electricity supply to rely on more renewables. In 2017, 28.2\% of the electricity was supplied by non-fossil fuel sources.

Electricity generated from renewable sources, such as solar, wind and hydro is characterized by strong diurnal, synoptic and seasonal variability. In systems with high penetration levels, their variations can be smoothed out by geographical aggregation, inter-connecting transmission, storage, demand-side management or local conventional power balancing. Storage units may be charged when excess electricity is present, and discharged at a later time. Reservoir hydro can be considered as storage facilities with an uncontrollable natural inflow, which supplies approximately 20\% of today's annual power demand. Transmission balances the fluctuations by transporting electrical energy geographically from sources to sinks, instead of moving electricity in the temporal dimension.

The cost of such complex systems, together with temporal availability of renewable generators, operational constraints of transmission lines, hydro reservoir cascades and storage charge/discharge and their CO$_2$ emission intensities, calls for a model, with a sufficient level of detail in time and space. Furthermore, to secure the optimal system configuration, long term validated continuous high-resolution weather data, which both renewable supply and power demand rely on, is a necessity.

In the Chinese power sector context, there have been numerous studies on its future transformations, from various perspectives. Multi-region planning models have looked into the chronological transition of the supply side based on today's composition and policy targets, and consequently the transmission infrastructure needed \cite{cheng2015multi,zhang2018multi,zhang2018system,yi2016inter,guo2016multi}. Ref. \cite{zhang2017integrated} co-optimized the system by considering the power supply and transmission components at the same time. Ref. \cite{chen2014assessing} assessed the low-carbon effects of inter-connections, and ref. \cite{yang2018regional} looked into the CO$_2$ reduction benefits of integrating renewables to district heating. Ref. \cite{he2016switch} was the first, which studied the Chinese power sector decarbonization by including renewables, storage, hydro, and transmission in a detailed dispatch model. It gave an overview of an 80\% carbon reduction scenario in 2050 supported by extensive data and operation considerations. As for our own work, our previous paper \cite{LIU2018534} presented an electricity network with the 31 provinces linked by transmission, in a 100\% renewable penetration scenario, supported by heterogeneous wind and solar installation layouts. Both the heuristic and the optimization methods gave lower capacity cost as well as backup reduction.

This study aims to clarify the role of hydro power, storage and transmission under ambitious CO$_2$ emission reduction scenarios of future highly renewable Chinese power systems. We consider the range of weather conditions that affect wind, solar and hydro power generation as well as electricity demand with a single, consistent 38-year-long dataset. The model optimizes on an hourly scale, whose numerical results then allow insights into the spatial-temporal patterns and correlations of the system components. Methodologically, this study takes a linear programming approach to minimize the total annualized investment and operation cost of generation, storage and transmission.

Additionally, we include the following novel features: the first calibrated high resolution reservoir hydro time series in a long-term
power system investment model, a comparison of costs under two distinct grid expansion strategies, an examination of hydro and storage operations at high resolution and an exploration of the solution space for the case that transmission volume is constrained.

It is important to properly represent hydroelectricity for
China, because it makes up a large fraction of China's electricity
generation and it has the potential to offer significant flexibility to
a system largely dependent on variable renewable generation. Many have tried integrating hydro power with other renewables for off-grid remote sites, regional or continental long term investment planning. Limited by either data availability \cite{Thangavelu2015Sep} or computational power \cite{Li2016Apr}, their representation of hydroelectricity is restricted to a handful of reservoir stations with low temporal resolution \cite{Khan2018Jan}. These time series from the TSOs \cite{Schmidt2016Jan} usually are in several discontinuous periods, and cascade coordinations are masked. Here, we use global reanalysis data so the method can be applied broadly, take a cross-disciplinary approach reaching to the hydrological sciences and calibrate the daily inflow time series model with historical data from 41 Chinese hydro stations as the first application. Furthermore, the inflows to reservoirs that are part of a cascade are constrained by the operation of their upstream stations. The cascades are coordinated together with the renewable generators and storage units in the system investment optimization.

As for transmission expansion, the studies mentioned above mostly showed inter-connection requirement with fixed generator fleets. Co-optimization of generation and transmission is particularly important for China, because some of its best wind and solar resources are in the Northwest, far from load centers, which would require significant expansion of transmission capacity for integration. Even today, significant amounts of solar power are curtailed owing to missing grid capacity. Ref. \cite{zhang2017integrated} and, in the European context, ref. \cite{ schlachtberger2017benefits} optimized the generation and transmission jointly. In our study, we not only co-optimize the two, but also compare the two grid expansion strategies of HVDC and UHVAC in addition to the regional grids, from a system-wide economic perspective. We also experiment constraining total transmission volume similar to refs. \cite{ schlachtberger2017benefits,Horsch2017Jun} and show the consequent shift in the investment landscapes with comparison to the European case.

Moreover, in the interest of transparency and reproducibility, we have made all raw data and codes associated with the article freely available \cite{Liu2018Dec,liu_hailiang_2018_1471322}, so that other researchers can examine and build upon the results presented here.

In this paper, the model is described in Section \ref{model} and the data in Section \ref{data}. In Section \ref{results1}, we first show an overview of cost optimal scenarios under the CO$_2$ emission reduction sweep and their cost allocation. Then, hydro and storage are explored in the temporal dimension in Section \ref{results2}, and Section \ref{results3} considers possible transmission volume constraints. Section \ref{conclusion} concludes the paper.

\section{Model}
\label{model}

We study scenarios of a future Chinese electricity system using the one-node-per-province network presented in our previous paper~\cite{LIU2018534}, but here we replace the heuristics with a linear techno-economic optimization \cite{brown2018pypsa,HORSC$H_2$018207} of total annualized cost:%
\begin{equation}
\label{eq:objective}
\min_{G_{n,s},F_\ell,g_{n,s,t},f_{\ell,t}} \left( \sum_{n,s} c_{n,s} G_{n,s} + \sum_{n,s,t} o_{n,s} g_{n,s,t}  + \sum_{\ell} c_{\ell} F_{\ell} \right)~.
\end{equation}
The indices $n$ label the nodes of the system, which represent the provinces in China without Hong Kong, Macau and Taiwan, but including the direct-controlled municipalities, Beijing, Shanghai, Tianjin and Chongqing. The system costs are composed of fixed annualised costs $c_{n,s}$ for generation and storage capacities $G_{n,s}$, variable costs $o_{n,s}$ for generation $g_{n,s,t}$, and fixed annualised costs $c_{\ell}$ for transmission capacity $F_{\ell}$ of line $ \ell $. The indices~$s$ label the generation and storage technologies comprising onshore wind, offshore wind, solar PV, super-critical coal power plants, open cycle gas turbines (OCGT), hydrogen storage (electrolysis and fuel cells for conversion, steel tanks for storage), central batteries (lithium ion) and reservoir hydro  generation. 

In all scenarios, only one year is solved at a time, due to computational limitations. The linear programming model was solved with the widely used commercial optimization software Gurobi \cite{gurobi}, specifically the logarithmic barrier algorithm. The model typically solves in 1–2 h per scenario on a Slurm-managed computer cluster using 4 cores (Intel \textsuperscript{\textregistered} Xeon \textsuperscript{\textregistered} Processor E5-2680 v3) and less than 35 GB of memory. This provides solutions whose accuracy can be measured by the closeness of the duality gap \cite{Arora2011Aug}, which in all simulations was at most \num{1e-6} of the total objective value.

The optimization has to satisfy a number of constraints listed in \tref{LP_model} and described in the following.

\begin{table*}[h!]
	
	\caption{Lists of decision variables, technical constraints and global constraints.}
	\label{LP_model}
	
	\begin{tabular}{lll}
		\toprule
		Decision variables &               Technical constraints &                     Global constraints \\
		\midrule
		generator capacities $ G_{n,s}$ &               power balance (Eq. \ref{powerbalance}) &  transmission volume limit $ CAP_{LV}$ \\
		storage capacities $E_{n,s}$ &      generation limit (Eq. \ref{eq:generation} and \ref{eq:availability}) &     CO$_2$ emission limit $CAP_{CO_2}$ \\
		line capacities $F_\ell$ &   renewable potential limit (Eq. \ref{eq:instpot}) &                                        \\
		&     storage operation (Eq. 6 and 7) &                                        \\
		&  hydro cascade operation (Sec. \ref{hydromodel}) &                                        \\
		&  transmission thermal limit (Eq. \ref{thermallimit}) &                                        \\
		&             grid topology (Table \ref{grid_topo_table}) &                                        \\
		\bottomrule
	\end{tabular}
\end{table*}

\subsection{Power balance}

To ensure a stable operation of the network, energy demand and generation have to match in every hour in each node. If the inelastic demand at node $n$ and time $t$ is given by $d_{n,t}$ then
\begin{equation}
\label{powerbalance}
\sum_{s} g_{n,s,t} - d_{n,t} = \sum_{\ell} K_{n\ell} f_{\ell,t} \hspace{1cm} 
\end{equation}
where $K_{n\ell}$ is the incidence matrix of the network~\cite{bollobas1998modern} and $ f_{\ell,t} $ is the power flow in $ \ell $ at time $ t $. This means the mismatch at node $ n $ between local power supply and demand is balanced by importing and exporting through the transmission network.

\subsection{Generators}

The dispatch of conventional fuel powered generators is constrained by the capacity $G_{n,s}$
\begin{equation}
\label{eq:generation}
0 \leq g_{n,s,t} \leq G_{n,s}.
\end{equation}

The maximum producible energy per hour in each installed unit of the renewable generators depends on the their local weather conditions, which is expressed as an availability $\bar{g}_{n,s,t}$ per unit of its capacity:
\begin{equation}
\label{eq:availability}
0 \leq  g_{n,s,t} \leq \bar{g}_{n,s,t} G_{n,s}.
\end{equation}
Note that excess energy can always be curtailed, e.g., by pitch angle control of wind turbines or disconnecting PV plants.
Only reservoir hydro power plants can delay the dispatch of the natural inflow to some extent by utilizing the storage reservoir, which is explained in detail in Section~\ref{hydromodel}.

The installed capacity itself is also subject to optimization, with a maximum limit $G_{n,s}^{max}$ set by the geographic potential:
\begin{equation}
\label{eq:instpot}
0 \leq  G_{n,s} \leq  G_{n,s}^{max}
\end{equation}

The capacity $G_{n,s}$ and the final dispatch $g_{n,s,t}$ of each generator are determined in the optimization such that they respect the physical constraints, while minimizing the total cost.

\subsection{Storage operation}

The state-of-charge $soc_{n,s,t}$ of all storage units has to be consistent with the charging and discharging in each hour and less than the energy capacity
\begin{align}
soc_{n,s,t} & = soc_{n,s,t-1} + \eta_{1} g_{n,s,t,\mathit{charge}} - \eta_{2}^{-1} g_{n,s,t,\mathit{discharge}} \nonumber \\
&\qquad + g_{n,s,t,\mathit{inflow}} - g_{n,s,t,\mathit{spillage}} , \\
0 \leq  soc_{n,s,t} & \leq h_{s,max} \cdot G_{n,s}.
\end{align}
The efficiencies $\eta_1, \eta_2$ determine the losses during charging and discharging, respectively. %
These losses also imply that the storage is only charged when there is oversupply of power available in the system, and discharged when the generators can not produce enough power and the import options are not sufficient.
The state-of-charge is limited by the energy capacity $E_{n,s} = h_{s,max} \cdot G_{n,s}$. Here, $h_{s,max}$ is the fixed amount of time in which the storage unit can be fully charged or discharged at maximum power.

The state-of-charge is assumed to be cyclic, i.e., it is required to be equal in the first and the last hour of the simulation: $soc_{n,s,t=0} = soc_{n,s,t=T}$. This is reasonable when modelling a full year, due to the annual periodicity of demand and seasonal generation patterns, and allows efficient usage of the storage at the beginning of the modelled time range.

Here, we restrict to two different energy storage technologies: lithium-ion battery and hydrogen storage \cite{aneke2016energy}. The former stores electricity as chemical energy. Batteries can be built in different sizes with capacity ranging from less than 100 W to several megawatts. Their $ h_{battery,max} $ is set at 6 hours, charge/discharge efficiency is assumed to be 0.9/0.9, making the round trip efficiency of 0.81 \cite{ZHANG2017397}. Hydrogen storage's efficiency, on the other hand, is on the low side, assumed at 0.75 (electrolysis) / 0.58 (fuel cells),  overall 0.435 for a round trip \cite{KAVADIAS2017}. This is partially compensated by the low storage energy costs and its low loss over time. $ h_{H_2,max} $ is assumed to be one week, i.e. 168 hours. Hydrogen, furthermore, is looked upon as a next generation clean energy carrier \cite{MCPHERSON2018649}.

\subsection{Inter-connecting transmission}

The transmission lines between provinces are simplified as a transport
model with controllable dispatch (a coupled source and sink),
constrained by energy conservation at each node.

The absolute flows on these transmission lines cannot exceed the line capacities due to thermal limits:
\begin{equation}
\label{thermallimit}
|f_{\ell,t}| \leq F_{\ell}.
\end{equation}
The line capacities $F_{\ell}$ can be expanded by the model if it is cost-effective to do so. To satisfy n-1 security requirements, a safety margin of 33\% of the installed capacity can be used \cite{brown2016optimising}. This
can be emulated a posteriori by increasing the optimized NTCs by a factor of $f_{n-1}=(1-\mathit{margin})^{-1}=1.5$.

The lengths of the interconnecting transmission lines $l_\ell$ are set by the distance between the geometric centers of the provinces, so that some of the transmission within each province is also reflected in the optimization. An assumed factor of 25\% is added to the line lengths to account for the fact that transmission lines may not be placed completely straight due to land use restriction.

\subsection{Transmission and CO$_2$ emission constraints}

The sum of transmission line capacities multiplied by their lengths is
restricted by a cap $\mathrm{CAP}_{LV}$, in the unit of MWkm, which is varied in different simulations:
\begin{equation}
\sum_{\ell} l_\ell \cdot F_{\ell} \leq  \mathrm{CAP}_{LV} 
\end{equation}
Line capacities are weighted by their lengths because the length
increases the cost as well. Please note, the cap used in the simulations are total transmission volume, meaning the distribution of this volume is not restricted here. Its distribution in the network follows the power balance optimization, and at the same time respects this total volume cap.

CO$_2$ emissions are also limited by a cap $\mathrm{CAP}_{CO_{2}}$, implemented using the
specific emissions $e_{s}$ in CO$_2$-tonne-per-MWh of the fuel of generator type $s$
and the efficiency $\eta_{s}$ of the generator:
\begin{equation}
\sum_{n,s,t} \frac{1}{\eta_{s}} g_{n,s,t}\cdot e_{s} \leq  \mathrm{CAP}_{CO_{2}}  \hspace{1cm} \leftrightarrow \hspace{0.5cm} \mu_{CO_{2}} \label{eq:co2cap}
\end{equation}
This cap is varied under different simulations to satisfy emission reductions goals, compared to today's level of approximately 6 billion ton of CO$_2$ annually from the power sector \cite{he2016switch}. And for simplicity, we do not consider the CO$_2$ emissions in the manufacturing and construction process of the generators, storage and transmission, only the emissions from OCGT and coal combustion are taken into account here.
The KKT multiplier $\mu_{CO_{2}}$ indicates,  in an unconstrained market, the CO$_2$ price necessary to obtain this reduction in emissions, i.e. shadow prices.

\FloatBarrier

\section{Data}
\label{data}

\subsection{Renewable potential}

The expansion of renewable capacities of solar, onshore and offshore wind is limited by local geography. In the simulations, nodal renewable capacities are optimized to scale wind and solar generation up and down, and this expansion is capped with its local geographical potential $ G_{n,s}^{max} $.

Here, we use a simple installation density to calculate the potential limits. Specifically, onshore wind turbine spacing is assumed to be 10 $ MW/km^2 $ and 5\% of the provincial territory is available for installation due to land use considerations. The values for offshore turbines are 10 $ MW/km^2 $ and 10\%, respectively \cite{Islam2013May}. And a 50 $ m $ sea depth is used to calculate suitable offshore sites. This assumption is justified because even though offshore farms today are mostly built within a depth of 20 $ m $, offshore foundation engineering has been showing huge improvements in recent years and floating turbines have shown high viability. PV panel farms, however, are more restricted by land use limits, so we set a territory fraction of mere 0.2\% and a density of 150 $ MW/km^2 $ \cite{Green2016Jul}. To simplify the calculations, we do not take into account distributed PV on house rooftops, high-rise windows, and so on.

\subsection{Wind, solar and load time series}

Hourly wind and solar power time series are modeled based on the Renewable Energy Atlas \cite{andresen2015validation}, which was validated for Denmark, Germany, China and has been used to produce time series for Europe \cite{victoria2018using}, US \cite{becker2015renewable} and Australia. In a nutshell, we combine gridded reanalysis weather data from CFSR \cite{cisl_rda_ds094.1} with technical specifications for wind turbines and solar PVs. Hourly weather data such as 10 $ m $ above ground wind speed, surface solar radiation, temperature can be used to calculate wind and solar power generation time series for selected areas. The calculations are done, in short, by interpolating turbine power curves \cite{staffell2016using} or simulating radiation on inclined PV panels \cite{pfenninger2016long}. The time series calculation is explained in detail in our previous studies \cite{LIU2018534,liu2017howto}.

These time series $ \bar{g}_{n,s,t} $ are calculated per unit capacity, meaning $ \bar{g}_{n,s,t} G_{n,s} $ represents the maximum renewable power available at time $ t $. Also, we recognize that different geographical layout of wind and solar capacities would give different power generation due to weather condition variations. Here, we simplify their distribution by assuming a uniform spreading of the capacities over the top 40\% of raster cells in each province, i.e.  $ \bar{g}_{n,s,t} $  stay the same regardless of capacity expansion.

The load time series $ d_{n,t} $ try to resemble the 31-province electricity consumption patterns in 2050. The yearly average loads are predicted based on GDP per capita projections. And the patterns are mainly derived from regional, climatic and economic characteristics. The load fluctuations in a region usually show seasonal, intra-week and intra-day variations. The details are described in our previous paper \cite{LIU2018534}. Additionally, we implemented the degree day theory to take into account ambient temperature's impact on electric cooling demand.

\subsection{Hydroelectricity time series}
\label{hydromodel}

\begin{figure*}[h!]
	\centering
	\includegraphics[width=\linewidth]{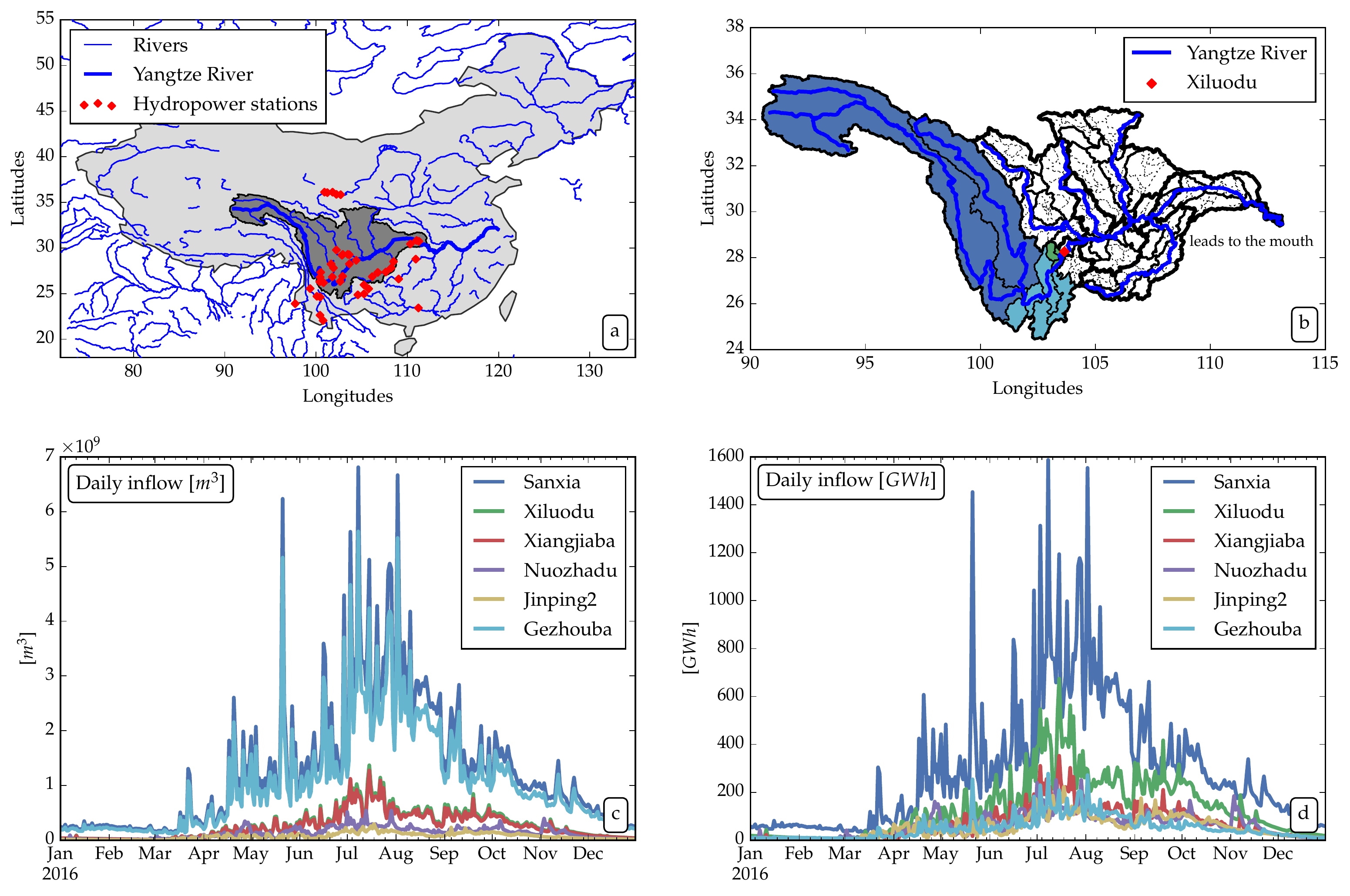}
	\caption{a. Forty-one largest hydro stations spanning major rivers in the southwest. Most dams are part of a cascade on the same river. The dark grey area represents the largest basin for Yangtze River. b. Upstream basins of hydro reservoirs are determined using the HydroBASINS dataset \cite{lehner2013global} and the Pfafstetter Coding System \cite{verdin1999topological}, the algorithm of which is described in detail in the appendix. This sub-figure shows the zoomed-in of the dark grey area in a. Thick solid lines represent higher level basin boundaries, and thin dashed lines enclose lower level basins. Hydro station \textit{Xiluodu} collects surface runoff from the colored areas, and different colors denote various basin levels. c\&d. Daily reservoir inflow time series in 2016 in terms of both water volume (left) and potential power production (right) for the 6 largest hydro stations. This dataset is freely available online at {http://doi.org/10.5281/zenodo.1471322} \cite{liu_hailiang_2018_1471322}.}
	\label{dams}
\end{figure*}

Here, we focused on 41 large-scale reservoir-based hydro stations in China, determined their corresponding basin areas, estimated their inflow based on gridded surface runoff data from CFSR \cite{cisl_rda_ds094.1} and calculated their daily inflow time series in terms of both flow volume and potential power generation. To our knowledge, no high resolution hydroelectricity generation time series have been modeled or validated before.

Electricity generation of both wind turbines and solar PV depends on local, instantaneous weather conditions. Wind speed and solar radiation do not necessarily affect the energy output at a location afar. For hydroelectricity, this is not the case. In fact, the vast majority of the water, whose potential energy is converted to electricity at the hydro station, is not from the raster cell it is in. We only consider the 41 largest hydro stations with a reservoir (\fref{dams}), i.e. run-of-river, whose generation varies upon instantaneous inflow is not included here.

Spanning over major rivers, hydro reservoirs' inflow is highly seasonal, and they depend on the precipitation in the upstream areas.  Usually, river basins are well-defined and documented from source to mouth. However, only basin areas which lie upstream of the hydro stations affect the reservoir inflows.

The HydroBASINS dataset \cite{lehner2013global}, is a series of polygon layers that depict basin boundaries and sub-basin delineations at a global scale. It provides a seamless global coverage of consistently sized and hierarchically nested sub-basins at different scales (from tens to millions of square kilometers), supported by the Pfafstetter coding scheme \cite{verdin1999topological} that allows for analysis such as up- and downstream connectivity. Basins, or watersheds were delineated in a consistent manner at different scales, and a hierarchical sub-basin breakdown was created following the topological concept of the Pfafstetter coding system.

Considering the fine spatial resolution in our model, we used basin levels 5, 6 and 7 of the HydroBASINS dataset. The larger the number, the finer the resolution is. Shown in Algorithm \ref{code_basin_determination}, we used the three most important features of the Pfafstetter scheme and determined the hydro stations upstream basins: Odd digits denote basin segments on the main stem, even digits denote tributaries of the main stem; at each level, higher digits denote upstream segments; a basin's Pfafstetter code with lower resolution is exactly the same as its finer scale basin taking out the last digit \cite{verdin1999topological}. An example is shown in \fref{dams}b, basins being separated by solid or dotted lines, upstream of hydro station \textit{Xiluodu} colored in green, teal and blue. Note that the rivers are only drawn to verify the basin delineations and they are not used in the upstream determination.

Surface runoff in the upstream basins are aggregated and calibrated against historical yearly reservoir inflow measurements \cite{Almanac}, to account for evaporation, transpiration, irrigation, groundwater infiltration or runoff movement. This time series is also made to account for the delay of runoff from upstream raster cells to the reservoirs, with an assumed flow speed of 1 $m/s$ \cite{yamazaki2009deriving}. The delays turn out to be ranging from 1 day to 2 weeks.

Finally, hydro stations' power production per unit water depends on their head heights. They are calculated by dividing the annual power generation by annual water inflow, and averaged over 7 years  \cite{Almanac}. Shown in \fref{dams}c and d, two pairs of hydro stations have identical inflow time series, due to their proximity over the same river, but their potential power generation are different from each other, for they are distinct in head heights.

One important character of hydro dams in China is that, they are usually part of a hydro station cascade, such as Three Gorges-Gezhouba, Xiluodu-Xiangjiaba,  Longyangxia-Laxiwa-Lijiaxia-Gongboxia-Qingtongxia \cite{SHANG201814,LU201556}. In such cascades, the dams are chained along the same river, and the downstream dams' inflow largely depends on their upstream stations' turbine control or spillage. This is also accounted for in the model optimization, assuming water flows into the downstream reservoir instantly. A number of time series for Three Gorges (Sanxia) and Gezhouba for weather year 2016 is shown in \fref{hydro_ts} in the later section.

The 41 selected hydro stations' annual generation explains over 80\% of the national hydroelectricity production, and in the model, their capacities are assumed to be fixed, even though a long list of major dams are currently in plan or under construction \cite{hess-20-3343-2016}. Indeed, the technically exploitable hydro power capacity is estimated to be 542 $GW$ \cite{gernaat2017high}, and only 341 $GW$ were put to use until 2017 \cite{huertas2017hydro}. The location of planned hydro stations may be known, but we are not able to validate the inflow or approximate the head heights until dam discharge measurement recorded. Therefore, the reader should bear in mind that hydro reservoir capacity in the future is expected to be significantly larger.

\subsection{Grid topology}

\begin{figure*}[h!]
	\centering
	\includegraphics[width=0.32\linewidth]{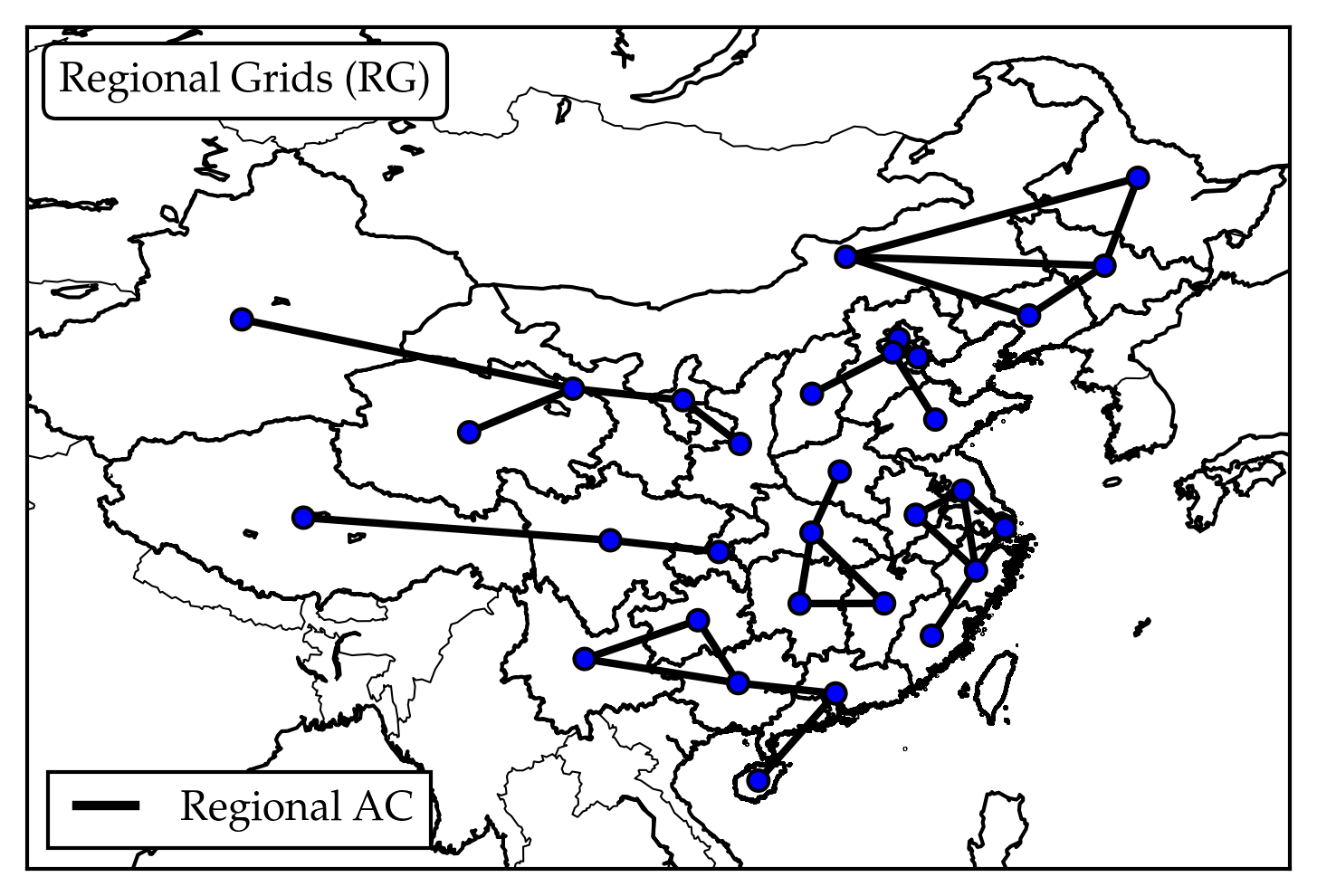}
	\includegraphics[width=0.32\linewidth]{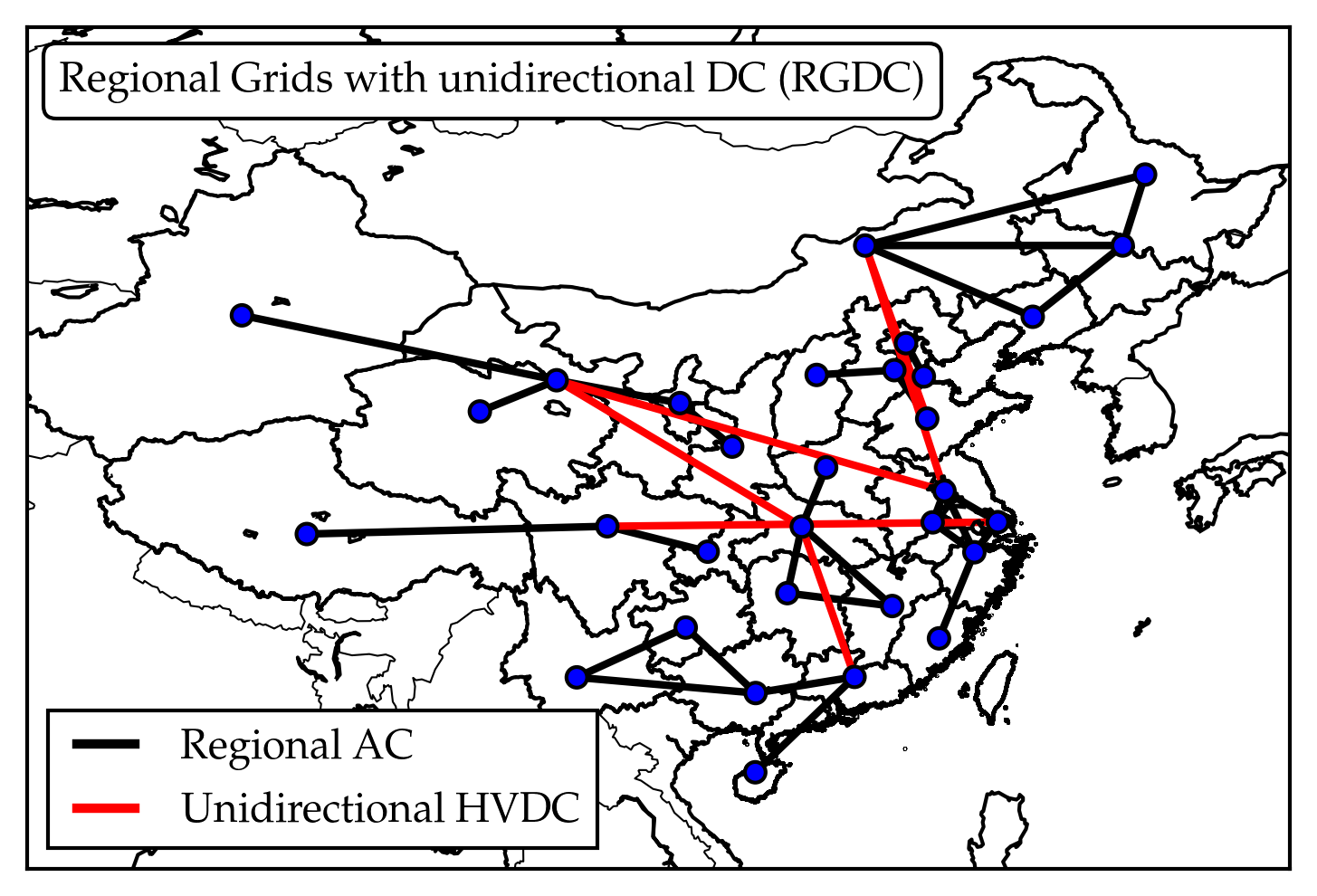}
	\includegraphics[width=0.32\linewidth]{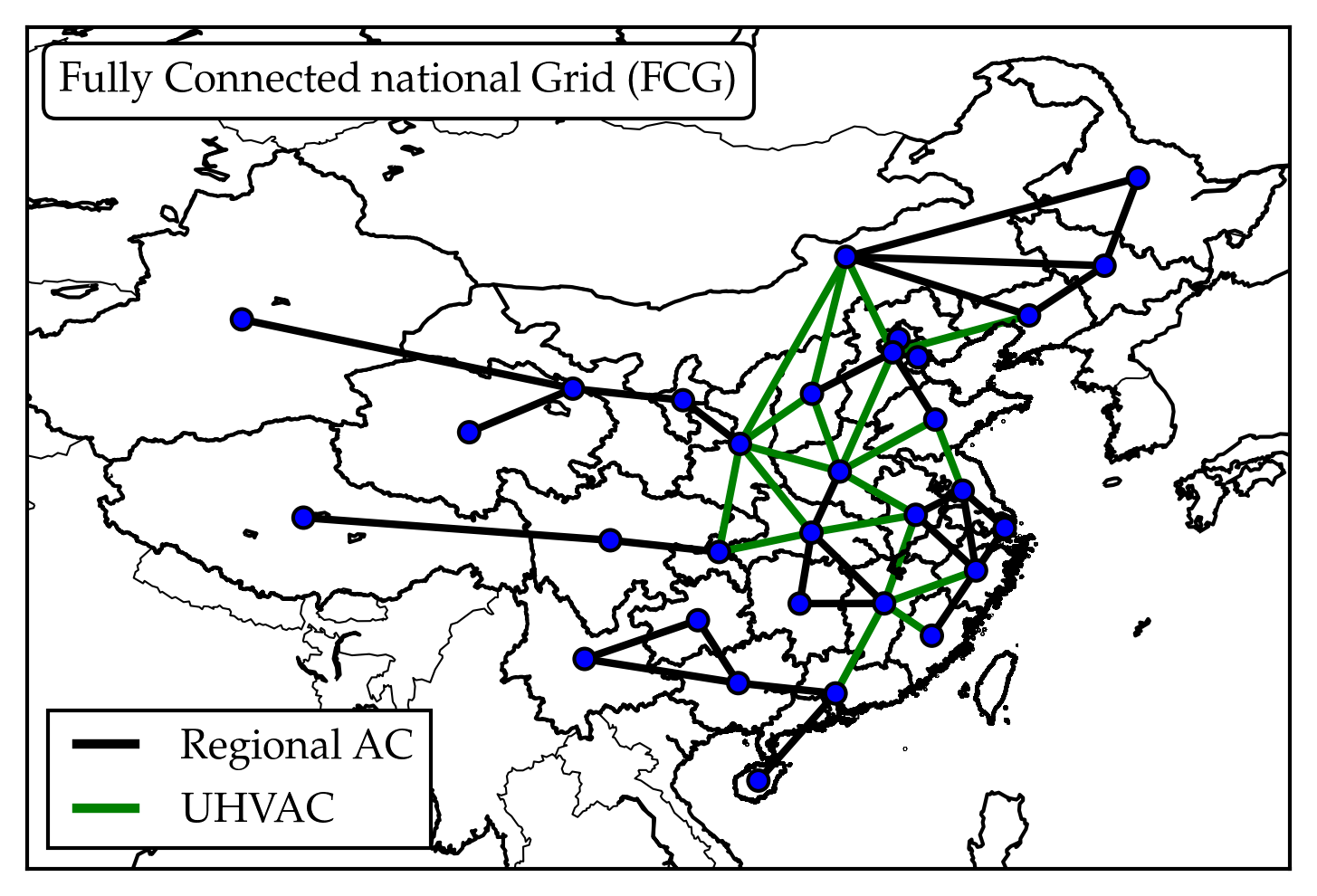}
	\caption{Grid networks: Regional Grids (RG), Regional Grids with unidirectional DC lines (RGDC) and Fully Connected national Grid (FCG). Note that the term "fully connected" here does not have the same meaning as in Graph Theory. The provinces are simplified as single nodes, located at their geometric center. In the middle graph, some nodes are slightly moved away from the centers to avoid confusion about the routes of the DC lines.}
	\label{grids_topo}
\end{figure*}

\begin{table*}[h!]
	\caption{The three grid topologies considered in the study.}
	\label{grid_topo_table}
	\begin{tabularx}{\textwidth}{ m{4cm} | X | X | X }
		\toprule
		This topology includes            & RG  & RGDC & FCG \\ 
		\midrule
		Regional AC         & Yes & Yes  & Yes \\
		Unidirectional HVDC & No  & Yes  & No  \\
		Bidirectional UHVAC & No  & No   & Yes \\ 
		\bottomrule
	\end{tabularx}
\end{table*}

\begin{table*}[h!]%
	\begin{threeparttable}
		\caption{Cost assumptions based on \cite{Schroder2013Current}, unless stated otherwise, which are calculated for 2030 using empirical curves. Particularly the values for wind are comparable to real construction projects in China reported by \cite{ouyang2014levelized}, while solar PV's cost projection is even lower. However, given the steep PV price drop in latest years, these low values are not far-fetched.}
		\label{tab:costsassumptions}
		\begin{tabularx}{\textwidth}{lrrrrrrr}
			\toprule
			
			Technology & \multicolumn{1}{l}{capital} & \multicolumn{1}{l}{fixed} & \multicolumn{1}{l}{marginal} & \multicolumn{1}{l}{lifetime} & \multicolumn{1}{l}{efficiency} & \multicolumn{1}{l}{cost per} & \multicolumn{1}{l}{$h_{max}$} \\
			& \multicolumn{1}{l}{} & \multicolumn{1}{l}{O\&M} & \multicolumn{1}{l}{cost} & \multicolumn{1}{l}{} & \multicolumn{1}{l}{} & \multicolumn{1}{l}{energy stored} & \multicolumn{1}{l}{} \\
			Units & \multicolumn{1}{l}{\euro/kW} & \multicolumn{1}{l}{\euro/kW/y} & \multicolumn{1}{l}{\euro/MWh} & \multicolumn{1}{l}{years} & \multicolumn{1}{l}{fraction} & \multicolumn{1}{l}{\euro/kWh} & \multicolumn{1}{l}{hour} \\
			
			\midrule
			
			onshore wind & 1182 & 35 & 0\tnote{a} & 25 & 1 &  &  \\
			offshore wind & 2506 & 80 & 0\tnote{a} & 25 & 1 &  &  \\
			solar PV & 600 & 25 & 0\tnote{a} & 25 & 1 &  &  \\
			OCGT\tnote{b} & 400 & 15 & 58.4\tnote{c} & 30 & 0.39 &  &  \\
			coal\tnote{b} & 1400 & 43 & 24.7\tnote{c} & 30 & 0.45 &  &  \\
			$ H_2 $ storage\tnote{d} & 737 & 12.2 & 0 & 20 & \multicolumn{1}{r}{$0.75,0.58$\tnote{e}} & 11.2 & 168 \\
			battery\tnote{d} & 411 & 12.3 & 0 & 20 & \multicolumn{1}{r}{$0.9,0.9$\tnote{e}} & 192 & 6 \\
			transmission\tnote{f} & {--} & 2\% & 0 & 40 & 1 &  &   \\
			hydro  & 2000\tnote{g} & 20 & 0 & 80 & 1 & N/A\tnote{g} & 2400\tnote{h}  \\
			\bottomrule
		\end{tabularx}
		
		\begin{tablenotes}
			\item [a] The order of curtailment is determined by assuming small marginal costs for renewables: 0.015 (onshore wind), 0.02 (offshore wind) and 0.01 (solar PV) \euro/MWh.
			\item [b] Open-cycle gas turbines have a CO$_2$ emission intensity of 0.19 t/MW$_{th}$, and supercritical coal-fired power plants 0.9 t/MW$_{th}$.
			\item [c] This includes fuel costs of 21.6 (gas) / 8.4 (coal) \euro/MWh$_{th}$.
			\item [d] Budischak et al. \cite{budischak2013cost}.
			\item [e] The storage round-trip efficiency consists of charging and discharging efficiencies $\eta_1 \cdot \eta_2$.
			\item [f] Interpolated from \cite{wangweb}. Ballpark values: 200 (regional AC), 200 ( long-distance HVDC) and 730 (UHVAC).  Unit: \euro/MW/km.
			\item [g] The installed facilities are not expanded in this model and are considered to be amortized. %
			\item [h] Assumed according to \cite{guntner2004simple}.
		\end{tablenotes}
	\end{threeparttable}
\end{table*}

Historically, the Chinese power grid has been run as seven independent regional grids, with each covering several geographically contiguous provinces as shown in \fref{grids_topo}a.  The grid companies were responsible for the power balance in their own region, supplied by coal primarily, and inter-regional power transfer was scarce \cite{bogdanov2016north}. In the model, the grids are represented as follows. Provinces are simplified as single nodes, located at their geometric center and connected by transmission lines.

From the 1990s, several large-scale hydro power stations have been built in the provinces marked in \fref{dams}a and subsequently long-range high-voltage direct current transmission lines have been erected to transport hydroelectricity to the eastern regions. In the new century, especially in the 2010s, wind and solar power installations grew particularly in the northwest. But their generation variability and low local demand caused high curtailment problems. Long-range transmission to the central-eastern provinces was the evident choice. The majority of these are built as ultra High-voltage (800 $ kV $) point-to-point unidirectional DC lines, as their primary purpose is to export renewable electricity. This way, the regional grid networks are linked by the DC lines, and at both ends, using converter stations direct current flows are converted from and to AC flows. This so-called RGDC topology (Regional Grids with unidirectional DC), shown in \fref{grids_topo}b, is simplified by combining today's DC lines and taking into account future construction plans \cite{lin2017cost}. The unidirectional DC added on top of the regional grids are, Gansu-Hubei, Gansu-Jiangsu, Inner Mongolia-Shandong, Inner Mongolia-Jiangsu, Sichuan-Shanghai and Hubei-Guangdong.

Another grid expansion strategy, that is heavily supported by the State Grid Corporation, is synchronizing the whole country by bidirectional Ultra High-Voltage (1000 $ kV $) AC networks, namely fully connecting the provinces as one national grid (FCG). Although its high cost and technical security are criticized by many, several UHVAC lines have been approved and under construction \cite{ming2016trans,zhang2018system}. Drawn in \fref{grids_topo}c, the simplified UHVAC lines connect the regional grids by multiple links, and a close-to-meshed grid in the east is formed.

The readers should note that the three grid networks are highly simplified, and do not represent reality to every detail. For instance, Inner Mongolia is actually covered by two separate grid companies in the eastern and western part of the province. And several HVDC lines are aggregated into one when they connect the same regional grids, for example, from Sichuan hydro stations to Shanghai or Jiangsu. Furthermore, the grid networks here only take the topology, but not their current transmission capacity into account, as in the future heavy capacity expansion is expected.

\subsection{Cost assumptions}

All cost assumptions are summarized in Table~\ref{tab:costsassumptions}. The given overnight capital costs were annualized with a discount rate of $7\%$ over the economic lifetime~\cite{zhao2016effectiveness}.

The transmission investment per line $\ell$ is calculated as:
$(C_t \euro/\si{MW}/\si{km} \cdot 1.25 l_\ell + c_{CP}) f_{n-1}$
with converter pair costs $c_{CP} = $ 150000 $\euro/\si{MW}$, if DC lines, and n-1 security factor $f_{n-1} = 1.5$. The unit cost $ C_t $ is interpolated with respect to length, from State Grid Corporation's budget reports \cite{wangweb} for AC, HVDC and UHVAC respectively.

The fixed operation and maintenance costs for transmission lines are 2\% of the investment cost~\cite{liu2015survey}.



\section{Results: A highly renewable China 2050}
\label{results1}

\subsection{Towards zero CO$_2$ emission}

\begin{figure*}[t!]
	\centering
	\includegraphics[width=\linewidth]{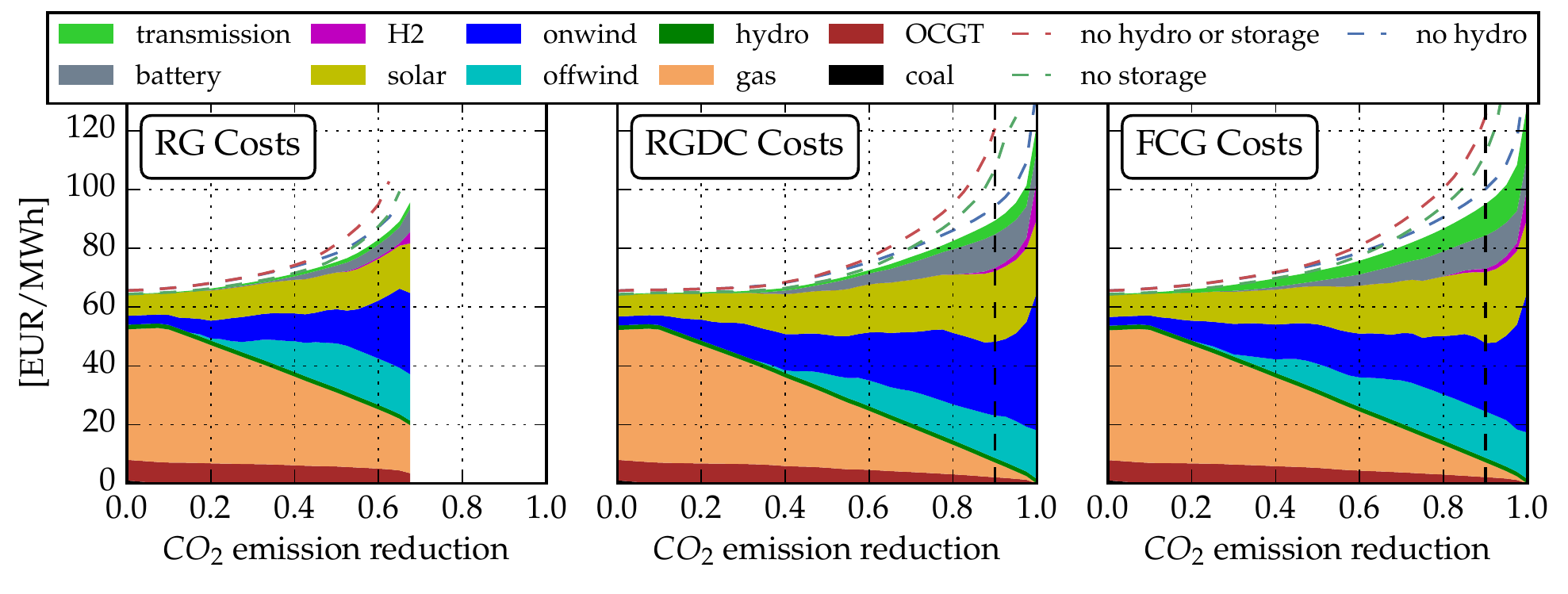}
	\caption{Component-wise average system costs on the pathway to 100\% CO$_2$ emission reduction under the three grid topologies, with a greenfield optimization. The dashed lines indicate average costs of systems with no storage (green), no hydro (blue) or neither (red). The black vertical dashed lines represent the 90\% reduction scenarios that we will be focusing on in later sections. Here, open-cycle gas turbines' costs are separated into OCGT (capital investments) and gas (marginal costs), as shown in the legend, the same goes for other figures.}
	\label{CR_cost}
\end{figure*}

\begin{figure}[t!]
	\centering
	\includegraphics[width=\linewidth]{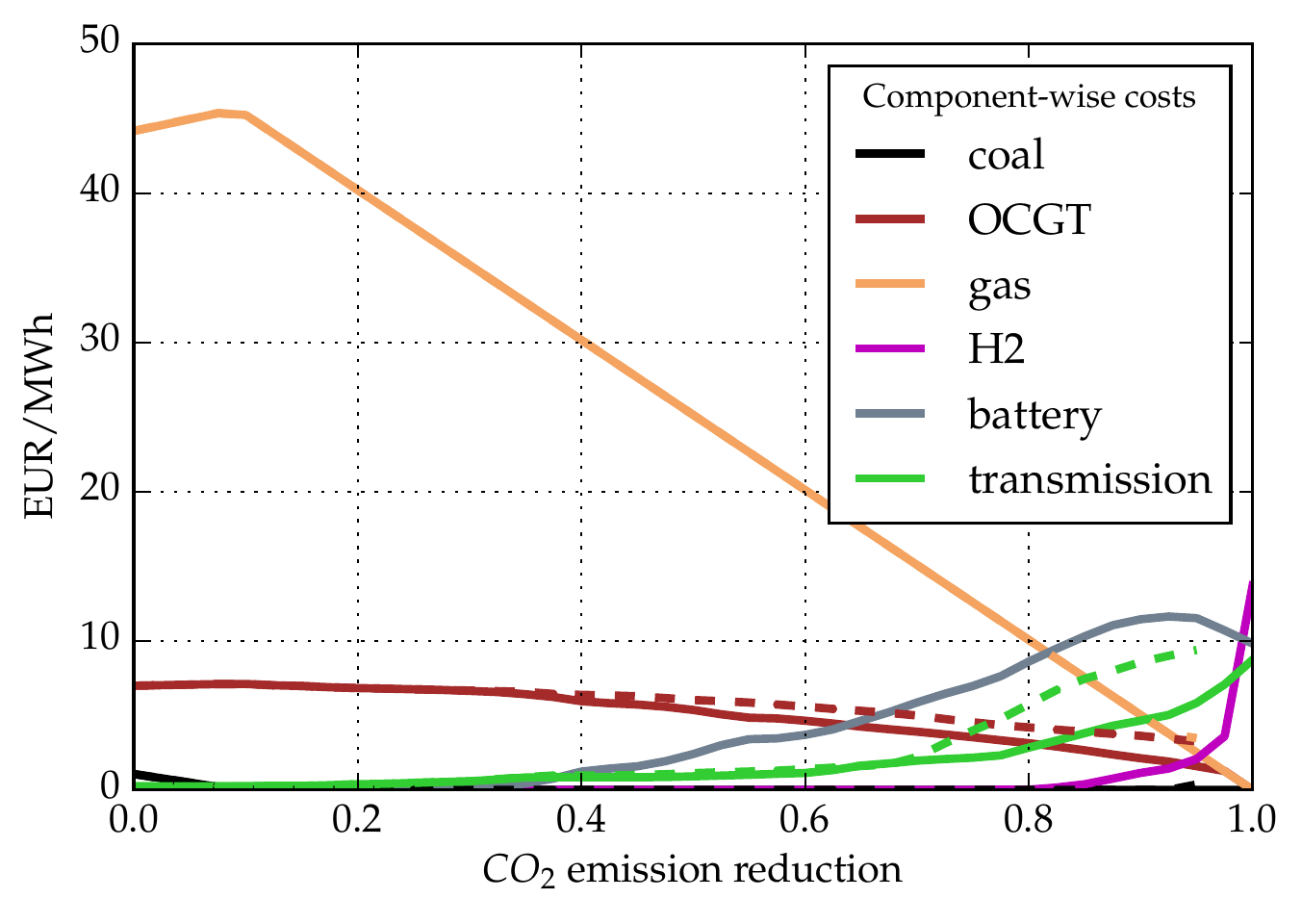}
	\caption{A selection of component costs of RGDC (with hydro and storage) on the pathway to 100\% CO$_2$ emission reduction, with a greenfield optimization. The dashed lines indicate the average component costs RGDC with hydro but no storage.}
	\label{unstacked_costs_RGDC}
\end{figure}

In this subsection, the feasibility of a Chinese power system with zero carbon emissions is examined using the weather year of 2016. The answer depends strongly on the transmission scenario.

The power sector pathway to 100\% CO$_2$ emission reduction, compared to today's value 6.0 billion metric ton per year \cite{he2016switch}, with the three grid topologies are shown in \fref{CR_cost}. In this study, we focus on a greenfield investment planning, meaning we ignore existing plants and infrastructure. This explains the low share of coal in the scenarios. We acknowledge that coal has been, and probably will be an important power source in China in the next ten years. At the same time, natural gas has emerged on the continent, playing a pivotal part in the energy sector reform, particularly the heating sector as of 2018 \cite{BibEntry2017Dec}. In this greenfield optimization, we focus on showing where their operational and cost characteristics put them on the decarbonization pathways.

On the left of \fref{CR_cost}, the traditional regional grids are able to reduce emission up to 67.5\% with hydro and storage. At this threshold, none of the regions with high demand, namely North, East, Central and South are able to supply enough renewable power, and the system becomes infeasible. The bottleneck lies at renewable generator installations, which are capped by their geographical potential. This limit, of course, is sensitive to the renewable potential assumptions we made, such as spacing and land use. We did a preliminary analysis on this, and it turns out that, a change of land use fraction to 10\%(onshore wind), 10\% (offshore wind) and 0.8\% (solar PV) \cite{schlachtberger2017benefits}, can lift this feasibility threshold to 100\% with higher system cost. A detailed analysis may be interesting but is beyond the scope of this paper.

On the other hand, in low power demand regions, Northeast, Northwest and West, installations only count up to 46.9\%, 20.3\% and 18.4\% of their potential, respectively. It is clear that, absence of inter-connections restricts renewable energy exploitation in provinces with high capacity factors. This problem is similar to what we are facing today. Provinces with high wind or solar resources attracted major investments for installations, but the belated affiliated transmission infrastructure causes high curtailment up to 30\% \cite{zhang2016reducing,XU2018585,wang2018short}. Furthermore, in terms of average system costs, at 67.5\% emission reduction RG costs 93.6 \euro/\si{MWh}, which is 24.5\% and 19.0\% higher than RGDC and FCG, respectively. This can be attributed to the fact that, more renewable generators are required to supply the same amount of electricity if installed in provinces with lower capacity factors \cite{LIU2018534}.

\begin{figure*}[t!]
	\centering
	\includegraphics[width=0.49\linewidth]{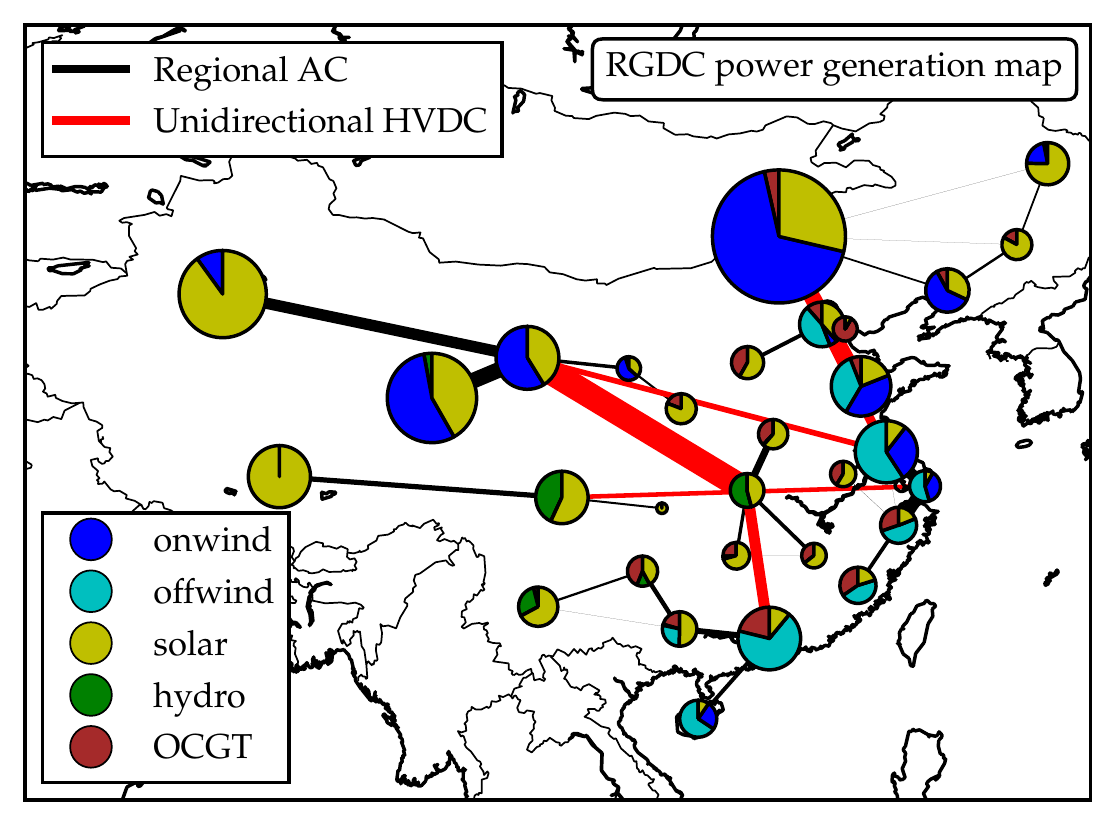}
	\includegraphics[width=0.49\linewidth]{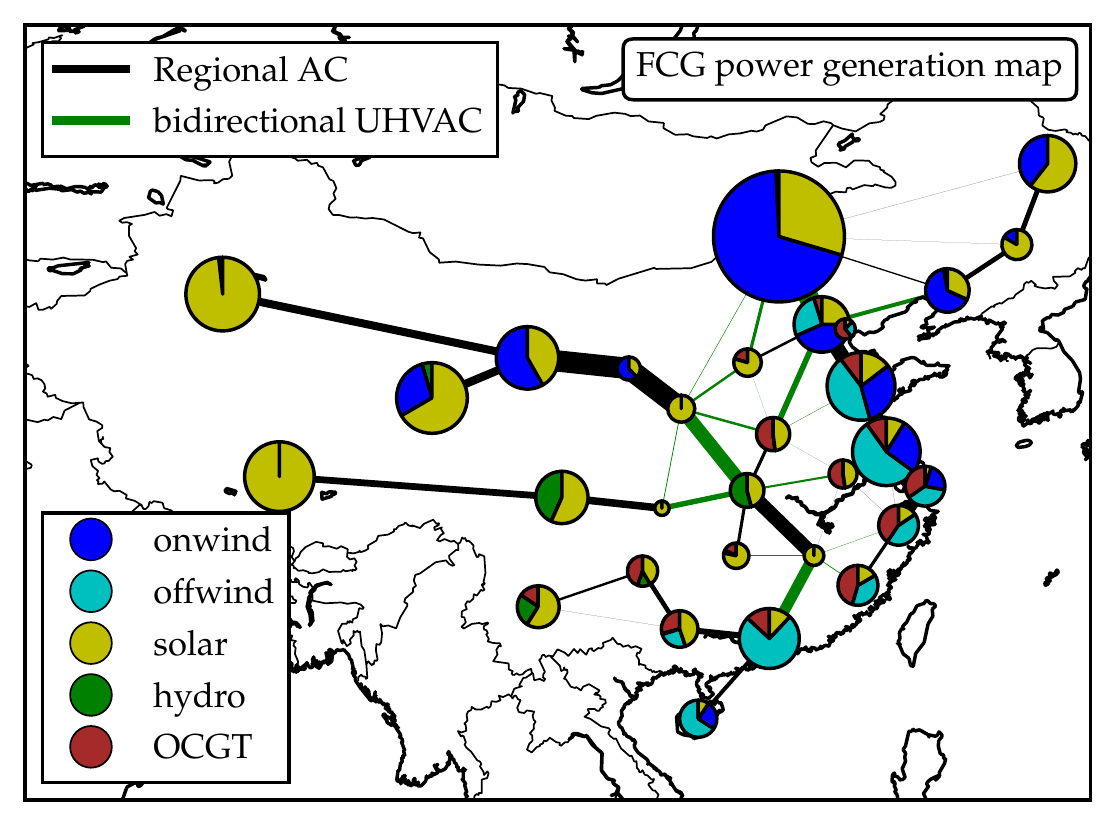}
	\caption{Average hourly power production at each node from the generation components onshore/offshore wind turbines, solar PV, reservoir hydro and OCGT, with 90\% emission reduction. For scale, the pie for Inner Mongolia (RGDC) measures 393 \si{GWh/h}, and the thickest HVDC (RGDC) 365.7 $ GW $.}
	\label{energy_map}
\end{figure*}

\begin{figure*}[t!]
	\centering
	\includegraphics[width=0.49\linewidth]{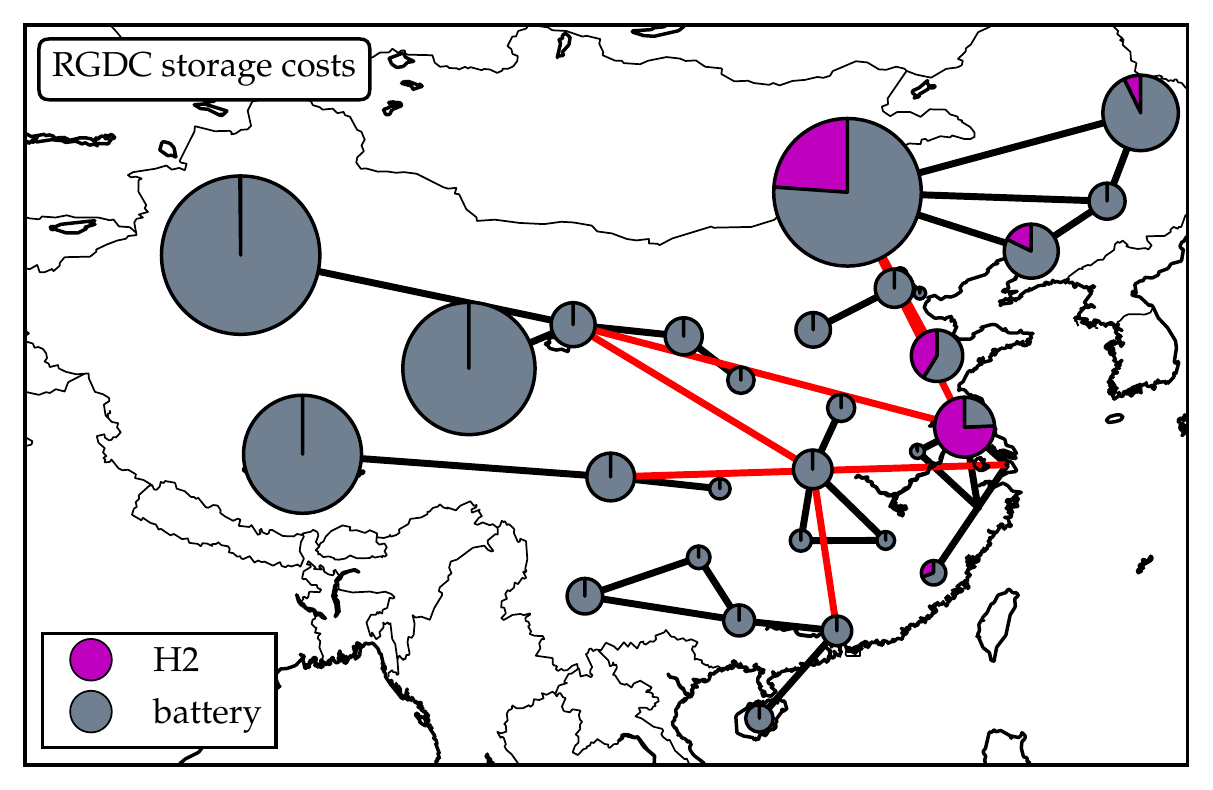}
	\includegraphics[width=0.49\linewidth]{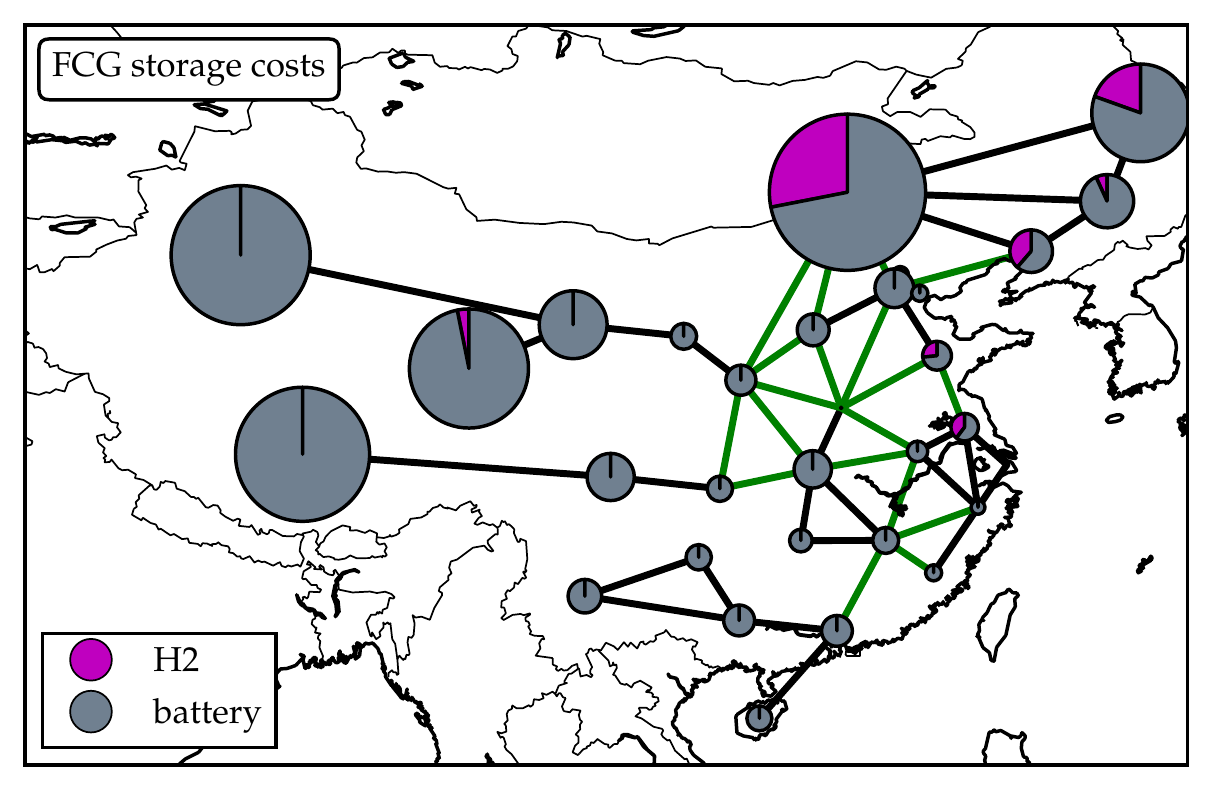}
	\caption{Costs of storage components battery and $H_2$, with 90\% emission reduction. For scale, the pie for Inner Mongolia (RGDC) measures 35.5 billion EUR, or 265 GW in storage capacity.}
	\label{storage_map}
\end{figure*}

As for RGDC and FCG systems, when the CO$_2$ emission reduction is smaller than 40\%, the total average cost is almost flat. Restricted gas combustion is replaced mainly by increased capacity of onshore wind and solar PV. None of transmission, battery or $H_2$ storage plays an important part under such scenarios. This indicates that 40\% emission reduction can be reached even with minimal interconnection and storage balancing units. This can also be deducted from the small differences among the dashed lines in \fref{CR_cost}. In \fref{unstacked_costs_RGDC}, for RGDC scenarios, we can see the interaction between gas-powered OCGT and supercritical coal power plants. Emission reduction targets below 10\% allow a small amount of coal-fired plants for their cheaper fuel costs, while OCGT prevails with increasing reduction limits. The almost flat OCGT capital costs up to 40\% emission reduction also implies that beyond this limit, gas-fired OCGT starts to play the role of balancing residual load. This can be seen as well, from the emergence of battery units in the system.

Another landmark reduction limit is around 70\%. In a RGDC network without storage, slightly increased OCGT capital cost (\fref{unstacked_costs_RGDC}) and renewable generators (not shown) can counter the effect of storage units, if reduction limit is below 70\%. Above this, significantly more transmission expansion---83.4\% more at 90\% emission reduction---is required, because increasing renewable generation alone is no longer the cost-optimal solution.

\subsection{Cost allocation}

\fref{energy_map} shows geographical distribution of power generation as well as transmission volume, while in \fref{storage_map} we map the component-wise storage costs, both for cost-optimal systems under RGDC and FCG, respectively. An immediate observation is that, renewable generators are  assigned predominately in the northwestern regions, and consequently provinces Inner Mongolia, Qinghai, Xinjiang, Gansu and Tibet supply 51.1\% (45.4\% for FCG) of the national load on average for RGDC. On the other hand, the geographically small provinces in the center and the east are mostly commissioned with insignificant amount of solar PV and OCGT, with the exception of coastal provinces of offshore wind installation (12.9\% of the national load). This is in good agreement to our previous study \cite{LIU2018534}, where we showed that in a cost-optimal system design, higher renewable capacity factor in the northwest shifts generator installations away from the demand center in the east.

Under the two grid networks, an almost identical distribution is observed for storage units. Battery and $H_2$ (\fref{storage_map}) follows the solar and wind installations, for their charge/discharge cycles match the respective generators' diurnal and synoptic fluctuations, respectively. In terms of cost, the five northwestern provinces take up almost 72\% of the national storage infrastructure.

This situation is quite different from that of Europe \cite{tranberg2018flow}, where major countries, such as Germany, Britain, Spain, France and Italy with overwhelmingly high power demand, are also blessed with decent wind or solar capacity factors as well as large installation potentials. Their geographically centrality on the continent makes it preferable to inter-connect them and the small countries nearby, forming a European transmission network. With these bidirectional links, the cooperation among the countries helps reduce each other's power mismatch and backup infrastructure, with moderate transmission expansion \cite{rodriguez2015localized,schlachtberger2017benefits}.

\begin{figure*}[t!]
	\centering
	\includegraphics[width=\linewidth]{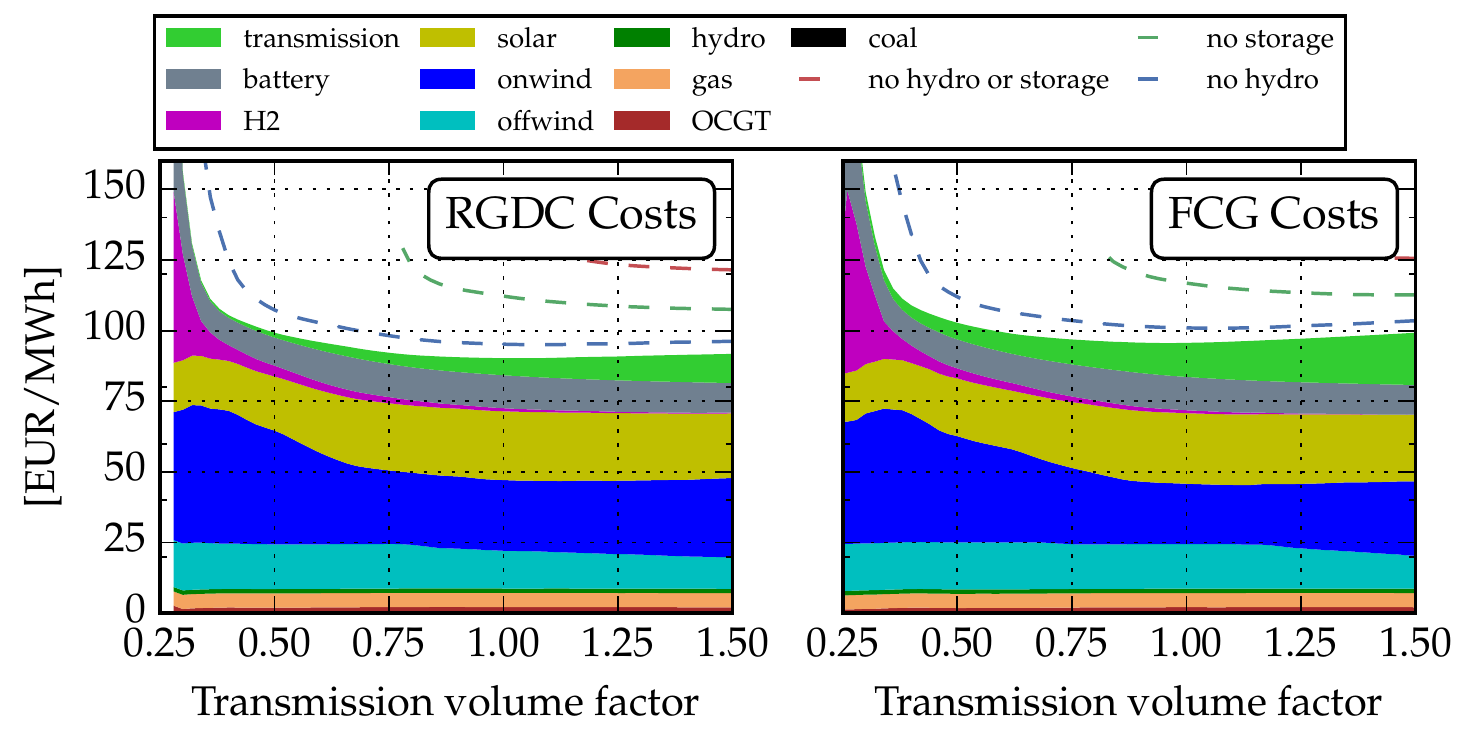}
	\caption{Component-wise average system costs plotted as a function of constrained transmission volume with respect to their optimal values 2231 and 1851 TWkm for RGDC and FCG, respectively in 90\% emission reduction scenarios, both with hydro and storage, in a greenfield optimization. The dashed lines indicate total average system cost with no storage (green), no hydro (blue) or neither (red).}
	\label{transmission_volume_plots}
\end{figure*}

The FCG grid expansion strategy seems to resemble the European case. However, our cost optimal analysis suggests UHVAC networks are not necessary, and unidirectional DC lines on top of the regional grids are sufficient economically as well. Marked as vertical dash-dot lines in \fref{CR_cost}, at 90\% CO$_2$ emission reduction, RGDC's average component-wise costs of renewable generators (onshore wind 24.8 \euro/\si{MWh}, offshore wind 23.8 \euro/\si{MWh}, solar PV 14.3 \euro/\si{MWh}) are all within 10\% variation of those for FCG. The same goes for storage units. The only significant difference is transmission cost 4.5 \euro/\si{MWh} and 10.5 \euro/\si{MWh} respectively. This can be attributed to the unit cost of UHVAC and FCG's high number of lines. However, if not provided any balancing units such as hydro or storage, FCG takes precedence when emission reduction goes higher than 90\% or transmission volume expansion is limited, which is shown in \fref{transmission_volume_plots}.

\begin{figure}
	\centering
	\includegraphics[width=\linewidth]{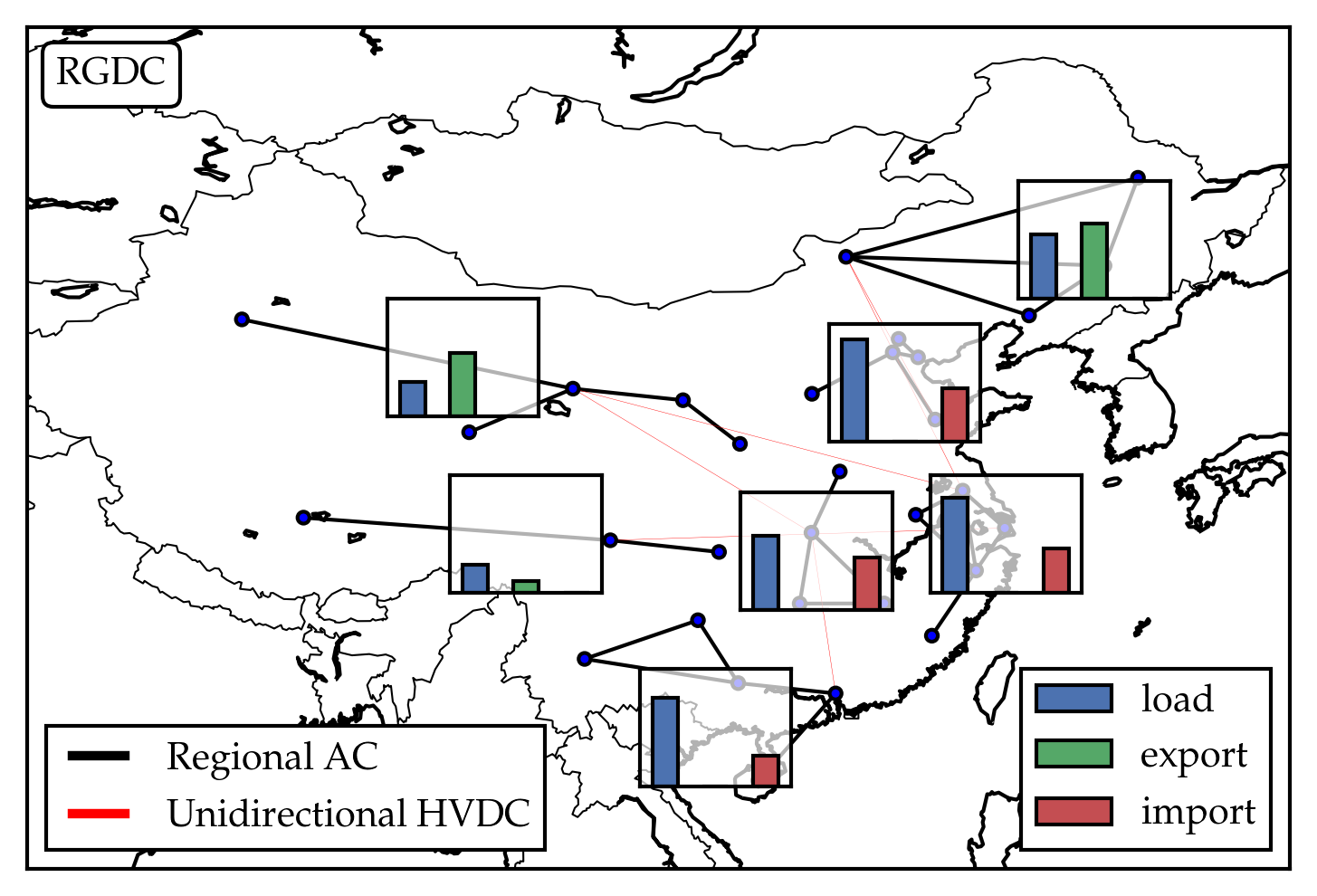}
	\caption{Hourly average local load, power export and import of the 7 regional grids under the 90\% emission reduction scenario, calculated based on the power flows on the unidirectional HVDC lines shown in red.}
	\label{RGDC_load_import_export}
\end{figure}

As for RGDC, its addition of DC connections are characterized by long length and going only in one direction. The diametric distribution of demand and renewable resources makes this unidirectional, long-range grid expansion more economically preferable than a fully connected national grid. The DC links also allow more insight into the cooperation among the regions. In \fref{RGDC_load_import_export}, with respect to their average load, Northeast and Northwest export 116\% and 186\% to other regions, while Central and East have to rely on imports for 70.7\% and 52.2\%, respectively.

\section{Results: Hydro and storage}
\label{results2}

In the optimal RGDC systems storage components cost 1.11, 1.55 and 11.3 \euro/\si{MWh} for $H_2$, hydro and batteries, respectively, which almost count up to offshore sector's cost, with 90\% emission reduction. Their role in the power systems is explored in this section.

\begin{figure*}[t!]
	\centering
	\includegraphics[width=\linewidth]{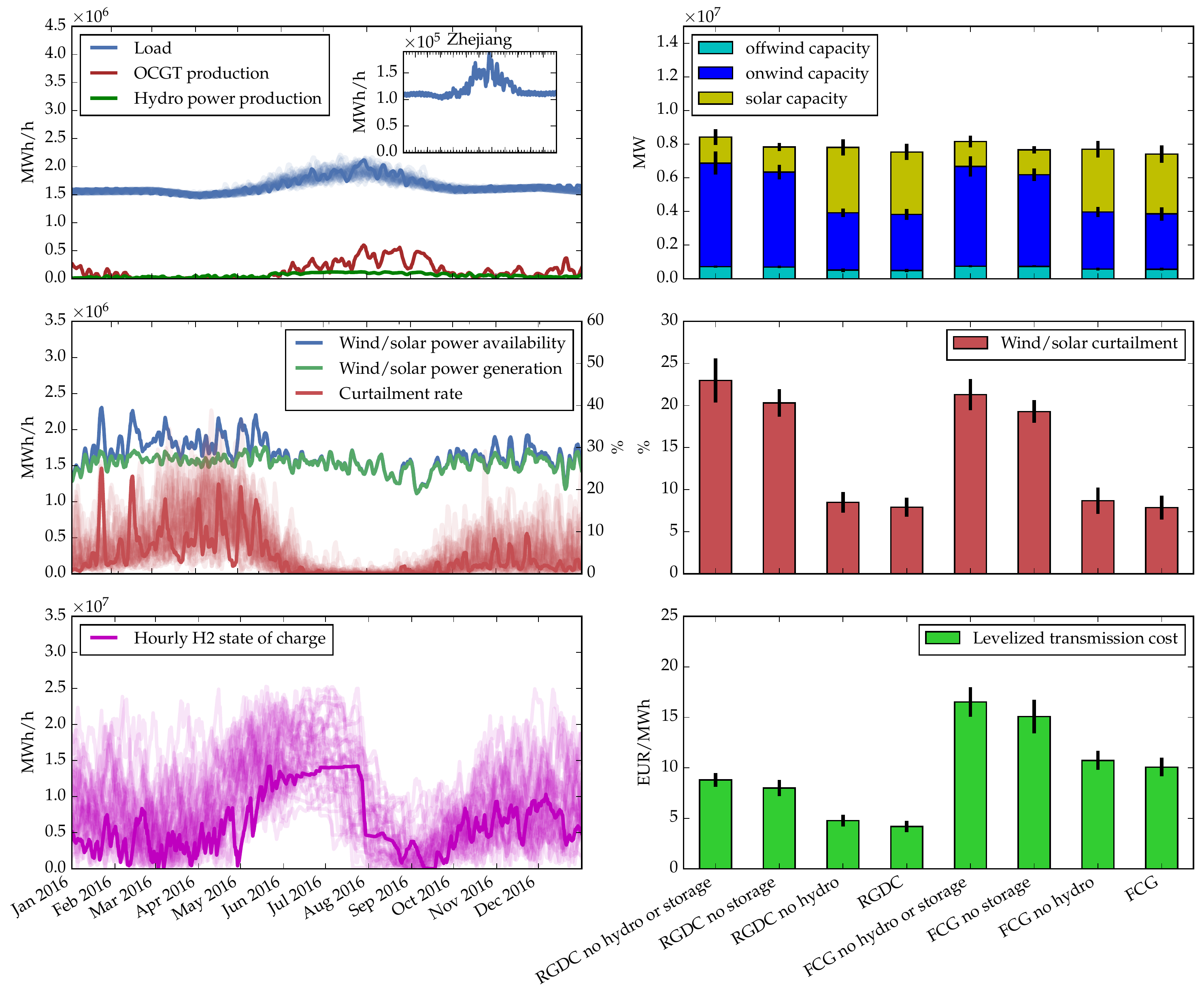}
	\caption{Time series on the left show the total load, OCGT and hydroelectricty production (top); wind and solar power availability/production, based on which curtailment rate (middle); state of charge of storage units $H_2$ (bottom). The solid lines represent weather year 2016, and shaded lines for years 1979-2015. Bars on the right illustrate the mean wind and solar installed capacities (top), curtailment (middle) and levelized transmission costs (bottom). The error bars indicate the standard deviations around the mean values for 1979-2016.}
	\label{balancing_time_series}
\end{figure*}

Spatio-temporal variations of wind speed, solar radiation, surface runoff and ambient temperature are all coupled together, consequently wind, solar and hydro generation as well as electricity load time series should not be analyzed separately \cite{kozarcanin2018climate}. The latter is especially important for China, whose heavy use of air-conditioning in the hot summer span primarily July and August. Considering the complexity of these systems, the results may strongly depend on the input weather data.  Here, we employed 38 individual years of reanalysis dataset from 1979 to 2016 \cite{cisl_rda_ds094.1}, used weather year 2016 as an example and plotted other years' results for reference in \fref{balancing_time_series}.

\begin{figure*}[t!]
	\centering
	\includegraphics[width=\linewidth]{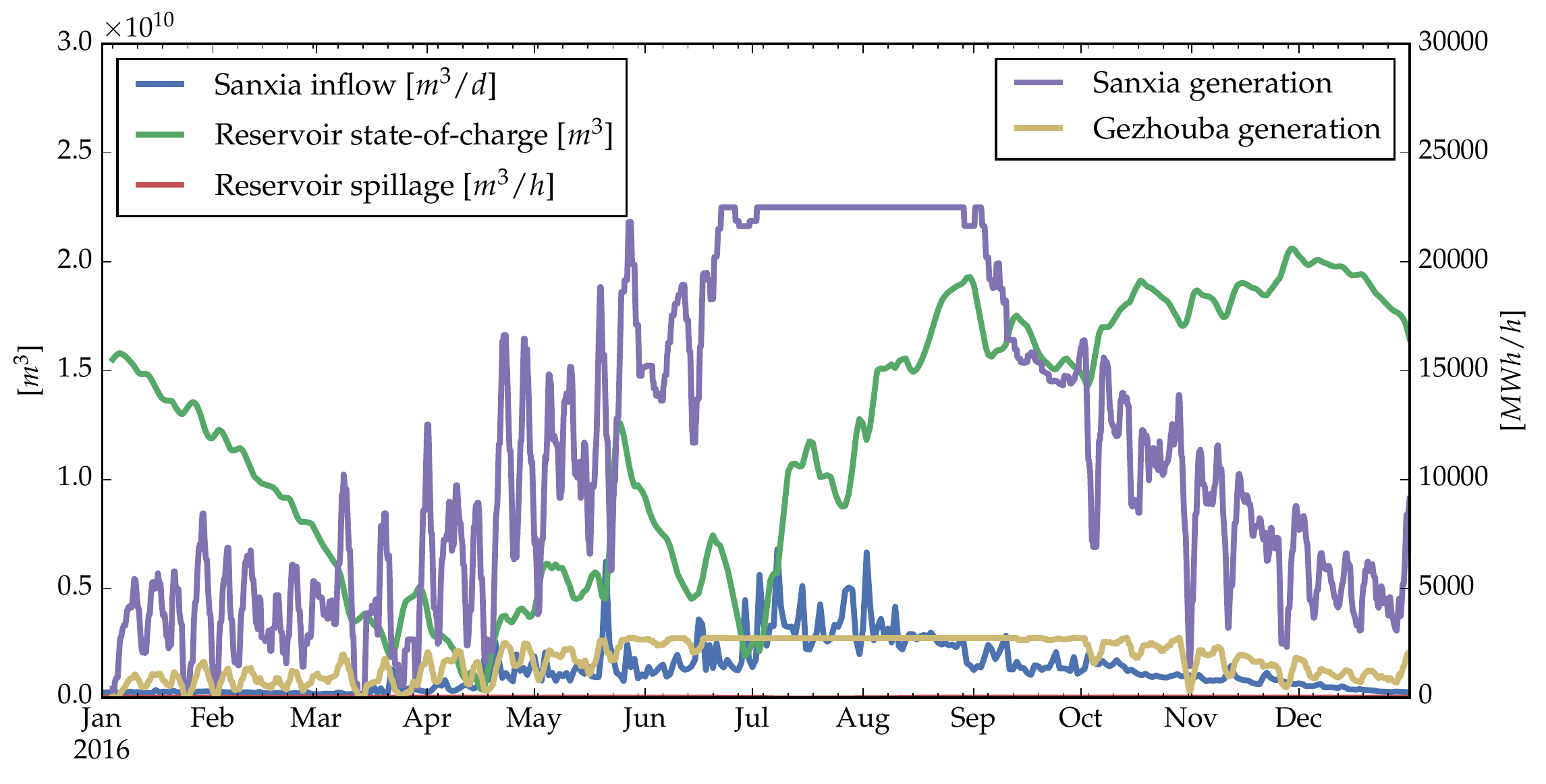}
	\caption{Visualization of reservoir inflow, state-of-charge, spillage, and power generation with weather year 2016 at \textit{Three Gorges (Sanxia)}, of which 38 km downstream lies another hydro station \textit{Gezhouba}. The former is equipped with a 39.3 billion $ m^3 $ reservoir, 181 m head height and generators totaling 22.5 GW, while the  latter 1.58 billion $ m^3 $, 47 m and 2.7 GW, respectively.  For better visualization, we only show 72-hour moving averages.}
	\label{hydro_ts}
\end{figure*}

\begin{figure}
	\centering
	\includegraphics[width=\linewidth]{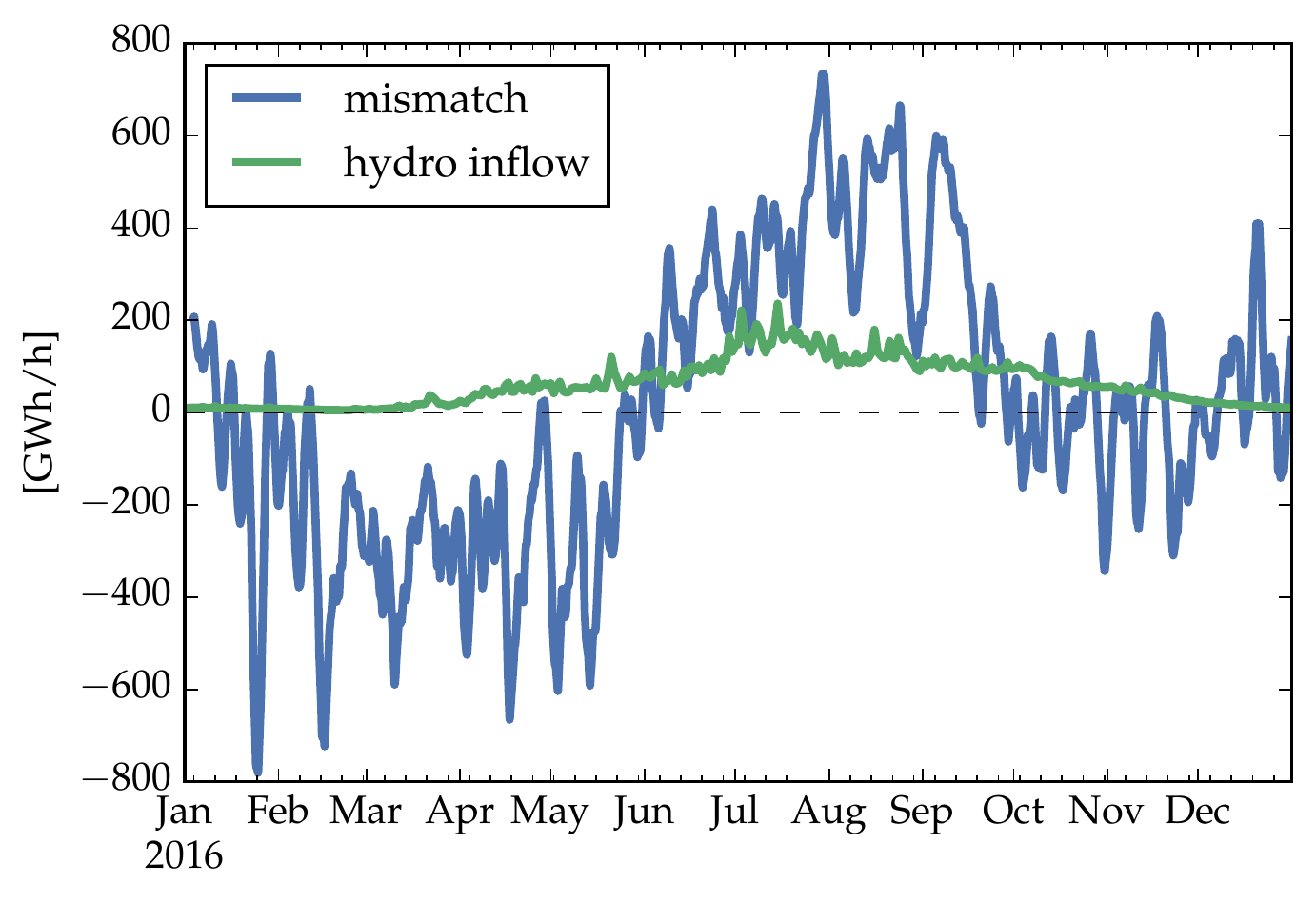}
	\caption{Mismatch between aggregated load and wind/solar power availability,
		shown as 72-h moving average, as well as total hydro reservoir inflow in terms of potential power generation.}
	\label{hydro_mismatch_time_series}
\end{figure}

One important character of the load for Chinese provinces is that, the fluctuation is flat throughout the year, with generally higher demand during summer, and peaks follow ambient temperature during heatwaves, which last several days. Zhejiang, for example, whose load time series is shown in the inset in \fref{balancing_time_series} has the highest 3-day average load at 187.2 GWh/h (July 25th), which is 55.6\% higher than the annual average and 85.7\% higher than the minimum (April 4th).

Renewable power availability, on the other hand, is lowest in summer and strong in spring, shown in \fref{balancing_time_series} (middle left). The time series fluctuation heavily depends on generator distribution evidently, but the seasonal trend is consistent. 
~In the first third of 2016, renewable power availability is 11.9\% higher than the second.

This diametric opposition in the temporal dimension between load and renewable supply, results in high curtailment in spring and heavy demand for long cycle $H_2$ storage, shown in \fref{balancing_time_series}. The 7.9\% average curtailmentfor RGDC (7.7\% for FCG), is already a reduction from 21.8\% (21.4\% FCG) for both networks with neither hydro power nor storage options, which is beneficial to generator investors and subsequently renewable expansion. Storage units, based on 38 single-year simulations, also help introduce more solar PV in the system and cut transmission costs by 54.5\% (44.2\% FCG). Battery storage's 6h charge cycle incorporates perfectly with solar PV's diurnal behavior, and consequently its capacity share overtakes onshore and offshore wind, compared to the scenarios with no storage (\fref{balancing_time_series} top right). Furthermore, halved inter-connection costs, shown in \fref{balancing_time_series} bottom right, indicates that locally storing renewable electricity for later use is more economically preferable than geographically spreading more renewable generators at the expense of higher transmission costs.

Hydroelectricity, which covers merely 3.6\% of 2050 annual load, can reduce average system costs by 6.2\% for RGDC (8.4\% FCG) at 90\% emission reduction. Even in systems absent of storage units, renewable generator capacity can be cut by 3.9\% (4.7\%), curtailment 7.1\% (7.6\%), transmission cost 18.2\% (13.0\%) (in \fref{balancing_time_series}). Its advantage lies in its seasonal inflow that is aligned with load and its reservoir-based storage-like flexibility.

The mismatch (load minus wind and solar availability), shown in \fref{hydro_mismatch_time_series}, is often negative in spring, and the excess renewable power from wind and solar generators can be stored in batteries or converted to hydrogen, otherwise curtailed. The peaks during summer are aligned with reservoir inflow, which, together with OCGT, act as an important supply and balancing sources. Looking deeper, hydro stations essentially act as storage units with fixed capacity and uncontrollable inflows. The operator may choose to store the water when there's a heatwave-induced demand peak forecast; or spill it downstream if there's heavy precipitation in its upstream area and the reservoir is almost full. For instance, 38 km downstream Three Gorges (Sanxia), there lies Gezhouba station, with a smaller reservoir and a smaller head.  \fref{hydro_ts} shows that, Sanxia has to keep the water level between its limits, coordinate with Gezhouba, minimize energy waste (spillage) and supply as much power as possible during summer. In fact, all hydro stations are running almost full load between June and September, shown in \fref{balancing_time_series} top left. 

The cost of limiting emissions can be analyzed with the help of the shadow price $ \mu_{CO_2} $. The shadow price is the dual variable of this constraint and can be interpreted as the price per unit of CO$_2$ the system is willing to pay to keep the emission below this cap. Its value increases from 228 to 293, 809 and 1307 $ EUR/tCO_2 $ (RGDC) if either or both of hydro and storage are striped.

However, something neither hydro nor storage options in the study are able to achieve is shifting the renewable electricity in spring to summer. This is shown in \fref{balancing_time_series}, where frequently up to 20\% of renewable power has to be curtailed and the longer cycle storage $H_2$, which is supposed to balance synoptic wind generation, is full during June and July and only discharges to meet a demand peak in August. The nearly zero curtailment also suggests that $H_2$ storage may not be used  in summer at all, since all renewable power must supply the load directly to avoid the non-unity efficiency of storage. This implies that longer-term energy storage is called for, and its realization has become a hot topic in the energy systems community \cite{REU2017290}.

\section{Results: Constraining transmission}
\label{results3}

Optimal inter-connection transmission volumes are 2231 and 1851 TWkm for RGDC and FCG, respectively in 90\% emission reduction scenarios. These values are 10 times higher than the European case found by similar studies \cite{schlachtberger2017benefits,SCHLACHTBERGER2018100}, which again can be attributed to the diametric mismatch of renewable power and load in this region. In this section, we will explore the possibility of reducing the transmission volume and consequently its impact on the interaction of the system components.

One evident observation from \fref{transmission_volume_plots}, shared by both grids, is that the solution space is quite flat while tuning the transmission factor by 25\% in both directions. This means, total system costs are to some extent insensitive to transmission infrastructure volume, which usually requires major investment from central TSOs or negotiations among major stakeholders. A 25\% transmission reduction can be replaced by an increase of onshore (from 22.9 to 24.4 EUR/MWh) and offshore (from 12.8 to 15.4 EUR/MWh) wind installations and a slight decrease of solar PV of 1.1 EUR/MWh for RGDC. The less volatile offshore wind prevails in this situation, since they are close to the high demand regions. And hydrogen storage also sees a 42.6\% sharp increase, for its charge cycle complements the synoptic behavior of wind generation. The cut in interconnections are primarily the more expensive DC lines, because transmission cost decreases by 41.1\% to 3.7 EUR/MWh, compared to the 25\% volume reduction. This reduces the country's energy dependence on the northwest, which could be beneficial from a grid security point of view.

Comparing to the European study \cite{schlachtberger2017benefits}, the authors found that even with a 95\% emission reduction target, European countries can archive electricity autonomy with zero interconnections at higher cost, solar is favored at lower transmission volume than wind and near zero storage is needed at the optimal solution with 285 TWkm transmission. China is not able to maintain power supply with a low interconnectivity, even for a fully connected grid (FCG). Furthermore, storage cost, especially battery, remains at relatively high level (10.8 EUR/MWh) despite a 150\% transmission volume compared to the optimal solution.

\FloatBarrier

\section{Discussions}

When it comes to limitations of our study here, we think it is necessary to address the question of linearization eligibility in the modelling framework, i.e. assuming linearity while representing the physical processes to an acceptable degree. We made major linear assumptions in reservoir hydro and transmission models. The head height of hydro stations, in reality, varies with current reservoir level, which depends on water inflow (calculated \textit{a priori}), reservoir shape, turbine control and spillage. Instead, the head is modeled as a constant and represented by power generation per unit water consumed on average in the years 2009-2015. The transmission is simplified as a transport model with controllable directed flows rather than modeling the node-by-node current and voltage \cite{Horsch2018May,brown2018pypsa}. And the power flows are approximated as linear DC power flows assuming small voltage-angle differences, low resistance and flat voltage profile, which are mostly true for long range high-voltage transmission \cite{Purchala2005Jun}.

\section{Conclusion and outlook}
\label{conclusion}

This study implements a detailed Chinese electricity network model and investigates future scenarios with high penetrations of renewable energy. It includes, among others, reservoir hydro, storage options, CO$_2$ emission reduction targets, transmission volume constraints as well as inter-connection network topology discussions.

We first compared the feasibility of going towards zero CO$_2$ emission for three different grid scenarios: one with regional grids only (RG), one with regional grids connected by HVDC lines (RGDC) and one with full UHVAC connections (FCG). It was shown that unidirectional HVDC lines on top of the regional grids can accommodate a highly renewable Chinese power sector, with the help of reservoir hydro, battery and $H_2$ storage. And we identified that in terms of CO$_2$ emission reductions compared to today's emissions, reducing below 40\% the system can sustain without these flexibility components, and above 70\% storage units become necessary to maintain reasonable system cost.

Allocating cost to the nodes, we showed that RGDC gave similar generator layout and storage infrastructure as FCG. This indicates that the diametric distribution of demand and renewable resources makes the unidirectionally transport renewable power to the east more economical, than UHVAC-connected "One-Net" national grid.  This, however, is merely a simplified techno-economical evaluation of the two, and is not able to account for antecedent system issues or demographical considerations.

We looked into the role of reservoir hydro and storage, by exploring the temporal dimension of the 90\% CO$_2$ reduction scenario. These flexibility components, can lower renewable curtailment by two thirds, allow higher solar PV share by a factor of two, decrease transmission cost significantly and also contribute to covering the summer peak demand.

Finally, constraining transmission volume by 25\% does not push total cost higher, but it demands more longer term storage and slightly increases renewable installation. The significant investment in short term battery storage remains even with copious transmission, unlike the European case.

When it comes to future work, coupling electricity to heating and transportation sectors is an evident choice to include more of China's emissions and to introduce more flexibility to the energy system. Electric vehicles may be a decentralized substitute for the demanding battery storage units, while power-to-gas can offer longer term storage to help shift the seasonal power mismatch.


\section*{Acknowledgments}

The authors thank Marta Victoria for fruitful discussions. The first author gratefully acknowledges the financial support from Idella Foundation Denmark and China Scholarship Council. T.B., D.P.S., G.B.A. and M.G. are fully or partially funded by the RE-INVEST project (Renewable Energy Investment Strategies -- A two-dimensional interconnectivity approach), which is supported by Innovation Fund Denmark (6154-00022B). T.B. also acknowledges funding from the Helmholtz Association, Germany under grant no. VH-NG-1352. The responsibility for the contents lies solely with the authors.

\section*{Bibliography}

\bibliographystyle{unsrtDOI}

\bibliography{hydro_paper}

\appendix

\onecolumn

\section*{Appendix}

\begin{framed}
	
	\nomenclature[01]{$n$}{nodes (provinces)}
	\nomenclature[02]{$t$}{hours of the year}
	\nomenclature[03]{$s$}{generation and storage technologies}
	\nomenclature[04]{$\ell$}{inter-connectors}%
	\nomenclature[05]{$c_{n,s}$}{fixed annualised generation and storage costs}
	\nomenclature[06]{$c_\ell$}{fixed annualised line costs}
	\nomenclature[07]{$o_{n,s}$}{variable generation costs}
	\nomenclature[10]{$e_{s}$}{absolute CO$_{2}$ emissions}%
	\nomenclature[11]{$d_{n,t}$}{demand}
	\nomenclature[12]{$g_{n,s,t}$}{generation and storage dispatch}
	\nomenclature[13]{$\bar{g}_{n,s,t}$}{availability per unit of capacity}
	\nomenclature[14]{$G_{n,s}$}{generation and storage capacity}
	\nomenclature[15]{$G_{n,s}^{max}$}{maximum installable capacity}%
	\nomenclature[20]{$f_{\ell,t}$}{power flow}
	\nomenclature[21]{$F_{\ell}$}{transmission capacity}
	\nomenclature[22]{$K_{n\ell}$}{incidence matrix of the network}
	\nomenclature[23]{$l_\ell$}{length of transmission line}
	\nomenclature[24]{O\&M}{operation and maintenance}%
	\nomenclature[25]{RGDC}{unidirectional DC on top of regional grids}
	\nomenclature[26]{FCG}{fully connected national grid}
	\nomenclature[27]{H$_2$}{molecular hydrogen}%
	\nomenclature[27]{HVDC}{high-voltage direct current}
	\nomenclature[27]{UHVAC}{ultra high-voltage alternating current}
	\nomenclature[27]{NTC}{net transfer capacity}%
	\printnomenclature
\end{framed}

\begin{algorithm*}
	\SetAlgoLined
	\DontPrintSemicolon
	\KwData{Three datasets consisting three levels of basins $ basins7, basins6, basins5 $}
	\KwResult{List of $ upstream\_basins $ that lie upstream of the hydro dam $ Dam $}
	$ coordinates \gets $ coordinates of $ Dam $\;
	\For{$ Basin7 \in basins7 $}{
		\If{$ coordinates $ is in $ Basin7 $}{
			$ PFcode  \gets pfcode(Basin7) $\;
		}	
	}
	append $PFcode$ to $upstream\_basins $\;
	\eIf{$ PFcode  $ is an even number, meaning it's a tributary}
	{ 
		Finish\;} (it's a main stem)
	{\For{$ Basin7 \in basins7 $}{
			$ p \gets pfcode(Basin7) $ \;
			{\If{$p > PFcode  $ and they are the same but last digit}
				{append $p$ to $upstream\_basins$\;}}	
		}
		$ PFcode \gets $ all but last digit of $ PFcode $\;
		\eIf{$PFcode$ is an even number, meaning it's a tributary}{Finish\;} (it's a main stem)
		{
			\For{$ Basin6 \in basins6 $}
			{$ p \gets pfcode(Basin6) $\;
				\If{$ p > PFcode  $ and they are the same but last digit}
				{append $p$ to $upstream\_basins$\;}
			}
			$ PFcode \gets $ all but last digit of $ PFcode $\;
			\eIf{$PFcode$ is an even number, meaning it's a tributary}{Finish\;} (it's a main stem)
			{
				\For{$ Basin5 \in basins5 $}
				{$ p \gets pfcode(Basin5) $\;
					\If{$ p > PFcode  $ and they are the same but last digit}
					{ append $p$ to $upstream\_basins$ \;}
				}
			}
		}
		
	}
	\caption{Determination of a dam's upstream basins}
	\label{code_basin_determination}
\end{algorithm*}

\end{document}